\documentclass[longauth]{aa} 

\usepackage{graphicx}
\usepackage{txfonts}
\usepackage{xcolor}
\usepackage{natbib}
\usepackage{multirow}
\usepackage{upgreek}
\usepackage{ gensymb }
\usepackage{float}
\usepackage{placeins}
\bibpunct{(}{)}{;}{a}{}{,} 

\begin{document}

\title{An ESPRESSO view of HD 189733 system.\thanks{Based on Guaranteed Time Observations (GTO) collected at the European Southern Observatory under ESO programme 1102.C-0744 by the ESPRESSO Consortium.}}

   \subtitle{Broadband transmission spectrum, differential rotation, and system architecture}

   \author{ E. Cristo\inst{\ref{inst1},\ref{inst2}}
   \and E. Esparza Borges\inst{\ref{inst3},\ref{inst4}}
   \and   N. C. Santos\inst{\ref{inst1},\ref{inst2}}
   \and   O. Demangeon\inst{\ref{inst1},\ref{inst2}}
   \and E. Palle \inst{\ref{inst4},\ref{inst21}}
   \and A. Psaridi \inst{\ref{inst19}}
   \and V. Bourrier \inst{\ref{inst19}}
   \and J.P. Faria \inst{\ref{inst1},\ref{inst2}}
   \and R. Allart \inst{\ref{inst24}}
   \and T. Azevedo Silva \inst{\ref{inst1},\ref{inst2}}
   \and F. Borsa\inst{\ref{inst5}}
   \and Y. Alibert \inst{\ref{inst23}}
   \and P. Figueira\inst{\ref{inst1},\ref{inst6}}
   \and J. I. Gonz\'alez Hern\'andez \inst{\ref{inst3},\ref{inst4}}
   \and M. Lendl \inst{\ref{inst19}}
   \and J. Lillo-Box \inst{\ref{inst9}}
   \and G. Lo Curto \inst{\ref{inst6}} 
   \and P. Di Marcantonio\inst{\ref{inst10}}
   \and C. J.A.P. Martins \inst{\ref{inst1},\ref{inst18}}
   \and N. J. Nunes \inst{\ref{inst14}}
   \and F. Pepe \inst{\ref{inst19}}
   \and J.V. Seidel \inst{\ref{inst6}}
   \and S. G. Sousa\inst{\ref{inst1}}
   \and A. Sozzetti\inst{\ref{inst8}}
   \and M. Stangret \inst{\ref{inst22},\ref{inst3},\ref{inst4}}
   \and A.~Su\'{a}rez~Mascare\~{n}o \inst{\ref{inst3},\ref{inst4}}
   \and H. M. Tabernero\inst{\ref{inst9}}
   \and M. R.  Zapatero Osorio\inst{\ref{inst9}}}

\institute{
 Instituto de Astrof\'isica e Ci\^encias do Espa\c{c}o, Universidade do Porto, CAUP, Rua das Estrelas, 4150-762 Porto, Portugal \label{inst1}
 \and
 Departamento de F\'isica e Astronomia, Faculdade de Ci\^encias, Universidade do Porto, Rua do Campo Alegre, 4169-007 Porto, Portugal \label{inst2}
         \and
 Instituto de Astrof\'{i}sica de Canarias (IAC), 38205 La Laguna, Tenerife, Spain \label{inst3}
 \and
 Universidad de La Laguna (ULL), Departamento de Astrof\'{i}sica, 38206 La Laguna, Tenerife, Spain\label{inst4}
\and
Centro de Astrobiolog\'\i a (CSIC-INTA), Crta. Ajalvir km 4, E-28850 Torrej\'on de Ardoz, Madrid, Spain \label{inst9}
 \and
European Southern Observatory, Alonso de C\'ordova 3107, Vitacura, Regi\'on Metropolitana, Chile\label{inst6}
\and 
INAF - Osservatorio Astronomico di Brera, Via Bianchi 46, 23807 Merate, Italy\label{inst5}
\and
INAF - Osservatorio Astronomico di Palermo, Piazza del Parlamento 1, 90134 Palermo, Italy\label{inst7}
\and
Département d’astronomie de l’Universit\'e de Gen\`eve, Chemin Pegasi 51, 1290 Versoix, Switzerland \label{inst19}
\and 
INAF - Osservatorio Astronomico di Trieste, via G. B. Tiepolo 11, I-34143 Trieste, Italy\label{inst10}
\and 
Consejo Superior de Investigaciones Cient\'{\i}cas, Spain\label{inst11}
\and
Department of Physics, and Institute for Research on Exoplanets, Universit\'e de Montr\'eal, Montr\'eal, H3T 1J4, Canada \label{inst20}
\and
Physics Institute, University of Bern, Sidlerstrasse 5, 3012 Bern, Switzerland\label{inst12}
\and
Instituto de Astrof\'isica e Ci\^encias do Espa\c{c}o, Faculdade de Ci\^encias da Universidade de Lisboa, Campo Grande, PT1749-016 Lisboa, Portugal \label{inst14}
\and 
Faculdade de Ci\^encias da Universidade de Lisboa (Departamento de F\'isica), Edif\'icio C8, 1749-016 Lisboa, Portugal \label{inst16}
\and
European Southern Observatory, Karl-Schwarzschild-Strasse 2, 85748  Garching b. M\"unchen, Germany \label{inst17}
\and
Centro de Astrof\'isica da Universidade do Porto, Rua das Estrelas, 4150-762 Porto, Portugal\label{inst18}
\and
Instituto de Astrofísica de Canarias (IAC), E-38200 La Laguna, Tenerife, Spain \label{inst21}
\and
INAF - Osservatorio Astrofisico di Torino, via Osservatorio 20, 10025 Pino Torinese, Italy \label{inst8}
\and
INAF - Osservatorio Astronomico di Padova, Vicolo dell’Osservatorio 5, 35122, Padova, Italy  \label{inst22}
\and
Center for Space and Habitability, University of Bern,  Gesellsschaftsstr. 6 CH3012, Bern\label{inst23}
\and
Department of Physics, and Trottier Institute for Research on Exoplanets, Université de Montréal, Montréal, H3T 1J4, Canada \label{inst24}
}
   \date{Received September 15, 1996; accepted March 16, 1997}

 
  \abstract
   {The development of state-of-the-art spectrographs has ushered in a new era in the detection and characterization of exoplanetary systems. The astrophysical community now has the ability to gain detailed insights into the composition of atmospheres of planets outside our solar system. In light of these advancements, several new methods have been developed to probe exoplanetary atmospheres using both broad-band and narrow-band techniques.}
   {Our objective is to utilize the high-resolution and precision capabilities of the ESPRESSO instrument to detect and measure the broad-band transmission spectrum of HD 189733b's atmosphere. Additionally, we aim to employ an improved Rossiter-McLaughlin model to derive properties related to the velocity fields of the stellar surface and to constrain the orbital architecture.}
   {The Rossiter-McLaughlin effect, which strongly depends on the planet's radius, offers a precise means of measurement. To this end, we divide the observation range of ESPRESSO into wavelength bins, enabling the computation of radial velocities as a function of wavelength. By employing a robust model of the Rossiter-McLaughlin effect, we first determine the system's color-independent properties across the entire spectral range of observations. Subsequently, we measure the planet's radius from the radial velocities obtained within each wavelength bin, allowing us to extract the exoplanet's transmission spectrum. Additionally, we employ a retrieval algorithm to fit the transmission spectrum and study the atmospheric properties.}
   {Our results demonstrate a high degree of precision in fitting the observed radial velocities during transit using the improved modeling of the Rossiter-McLaughlin effect. We tentatively detect the effect of differential rotation with a confidence level of $93.4 \%$ when considering a rotation period within the photometric literature values, and $99.6\%$ for a broader range of rotation periods. For the former, the amplitude of differential rotation ratio suggests an equatorial rotation period of $11.45\pm 0.09$ days and a polar period of $14.9\pm 2$. The addition of differential rotation breaks the latitudinal symmetry, enabling us to measure the true spin-orbit angle $ \psi \approx 13.6 \pm 6.9 ^\circ$ and the stellar inclination axis angle $ i_{\star} \approx 71.87 ^{+6.91^\circ}_{-5.55^\circ}$. Moreover, we determine a sub-solar amplitude of the convective blueshift velocity$V_{CB}$ $\approx$ $-211 ^{+69} _{-61}$ m$\,$s$ ^{-1}$, which falls within the expected range for a K-dwarf host star and is compatible with both runs. \par
   Finally, we successfully retrieved the transmission spectrum of HD 189733b from the high-resolution ESPRESSO data. We observe a significant decrease in radius with increasing wavelength, consistent with the phenomenon of super-Rayleigh scattering.}
   {}


   \maketitle
%
\section{Introduction}
The detection and characterization of exoplanets depend on several techniques that enable us to uncover the subtle signatures left by planets in the signals of their host star. One of the primary methods for detecting exoplanets is  radial velocities (RVs), which is used to measure the star's reflex motion around the system's barycenter and thereby detect planetary companions. Precisely measuring RVs is often challenging, and requires state-of-art, high-resolution spectrographs \citep[e.g.][]{Mayor2003, Pepe2021} with long-term stability.However,  RVs are not limited to detection alone; they are also valuable for the atmospheric characterization of exoplanetary systems, particularly during transits or occultations. \par
A transiting exoplanet covers stellar regions with varying brightness, spectral content and velocities. This is a result of the limb-darkening effect, the presence of stellar activity such as spots and planes, and the intrinsic rotation of the star. In photometric observations, this results in a decrease in the observed brightness of the star. When measuring RVs, it manifests as an anomaly caused by the planet blocking areas of the stellar surface with different projected velocities. This RV variation, first observed in binary stars, is known as the Rossiter-McLaughlin (RM) effect  \citep{Holt1893, Rossiter1924, McLaughlin1924}. The RM effect, for transiting exoplanets, was first measured for the HD 209458 system \citep{2000A&A...359L..13Q} and, more recently, successful attempts have been made to measure the RM anomaly within the solar system, such as the Earth \citep[e.g.][]{2015MNRAS.453.1684M,2015ApJ...806L..23Y} and Venus \citep[e.g.][]{10.1093/mnrasl/sls027}.\par
The RM effect can be used as a tool to complement the orbital geometry derived from the Keplerian motion outside transit, providing a direct way to measure the projected spin-orbit angle. Some notable applications of these orbital measurements include statistical studies about orbital tilts \citep[e.g.][]{2009ApJ...696.1230F, 2022PASP..134h2001A, 2022A&A...664A.162M}, and simultaneous measurements of the spin-orbit angles in multi-planetary systems \citep{2021A&A...654A.152B}. Additionally, since the planet covers different portions of the star along its track during a transit, the RM curve can also be explored to measure, e.g., the existence of differential motions of the stellar surface \citep[e.g][]{Cegla2016} or even to probe the spectra of the star behind the planet \citep{Dravins2021}.\par
Focusing on the exoplanet studies, RVs have become a source for the robust detection of atmospheres and the chemical species that compose them, using a variety of techniques \citep[e.g.][]{Sing2009, Wyttenbach2015, Nikolov2018, Ehrenreich2020, 2021A&A...646A.158T}, primarily during transits. When an exoplanet with an atmosphere transits, the radiation from its host star is filtered in the evening and morning terminators, encoding information about the composition and physical properties of the underlying atmospheric processes in the observed stellar spectrum. These properties are a function of atmospheric pressure and temperature. At high altitudes, where the pressure is lower, processes such as atmospheric dynamics, the presence of clouds, hazes, and temperature inversions dominate over the typical chemical reactions timescales \citep[e.g.][]{2014RSPTA.37230073M}. If we examine even higher altitudes, the intense radiation fields induce photochemical reactions that determine the atmospheric composition on these layers. In this region, at visible wavelengths, we frequently observe the signature of ionized alkali metals or molecules \citep[e.g.][]{2021MNRAS.505..435S, 2022A&A...666L..10A,2022MNRAS.513L..15S}.\par
\begin{table*}[ht!]
    \centering
    \caption{Observation summary of HD 189733b observations.}

    \begin{tabular}{cccccccc}
    \hline
    \hline
    &  Date &  N. Obs. & Int. Time [s] &  SNR@580 nm & Seeing (´´) & Airmass & $\sigma_{RV}$ [cm\,s $ ^{-1} $] \\
    \hline
    Night 1 & August 11, 2021 & 41 & 300 & 156 & 0.69-1.73 &1.48-2.08 & 32\\
    Night 2 & August 30, 2021 & 43 & 300 & 167 & 0.46-0.88 &1.48-2.01 & 32\\
    \hline
    \end{tabular}
    \label{tab:observations_summary}
    \tablefoot{The seeing and airmass ranges correspond to the minimum and maximum values during each night.}
\end{table*}
\begin{figure*}[ht!]
    \centering
    \includegraphics[width=\linewidth]{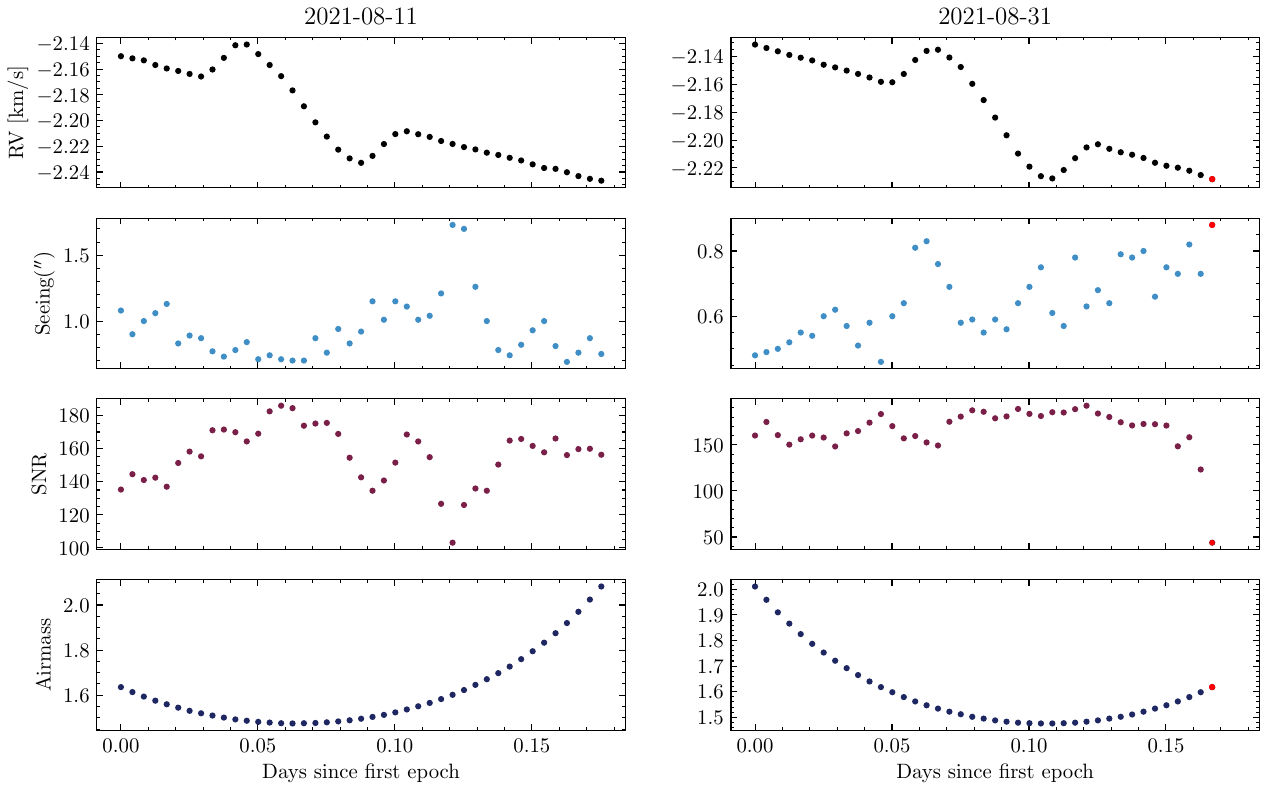}
    \caption{ Radial velocities of HD 189733b retrieved from the CCF header for the white light. The error bars are not visible since they are smaller than the dimension of the markers. In the same column, seeing evolution during the night, the signal-to-noise ratio at around $580$ nm and airmass as a function of the number of days since the first epoch. The red dot represents the observation that was removed, from the second-night set, due to low SNR.}
    \label{fig:header_data}
\end{figure*}
In this paper, we analyze the HD 189733 system using high-resolution ESPRESSO data. The star of this system is a K2V dwarf \citep{2003AJ....126.2048G} located at a distance of approximately $19.3$pc from Earth. It has a V-band magnitude of 7.6\footnote{The stellar atmospheric parameters and chemical abundances of HD189733 can be found in \citet{Sousa2018}.}  and belongs to the group of variable stars known as BY Draconis \citep{2010MNRAS.403.1949K}. The known planet orbiting this star, HD 189733b, was one of the first hot Jupiters discovered \citep{Bouchy2005} and has since been extensively studied due to its favorable planet-to-star radius ratio. It was the first exoplanet to have its surface temperature mapped \citep{2007Natur.447..183K}, and one of the first (along with HD 209458b) to have its atmosphere measured using spectroscopy with infra-red data from Spitzer \citep{2007ApJ...658L.115G}. From the visible to the IR, this planet is known for the characteristic decrease in radius with increasing wavelength. This phenomenon is thought to be the caused by the interaction of very small particles (smaller than the wavelength which is being observed) in the upper atmosphere, and it is commonly referred to as  Rayleigh scattering \citep[e.g.][]{Pont2008, Etangs2006}. The presence of this effect partially attenuates atomic and molecular signatures of the transmission spectrum.\par
Our goal in this work is to measure the broad-band transmission spectrum of HD 189733b using the high-resolution capabilities of ESPRESSO. To achieve this, we will use the chromatic Rossiter-McLaughlin \citep[CRM,][]{DiGLoria2015} method implemented in \texttt{CaRM} \citep{Santos2020, 2022A&A...660A..52C}.
This approach for retrieving transmission spectra was first used by \citet{Snellen2004}, who attempted to measure an increase in RM amplitude near the sodium D lines.\par

In section 2, we provide a summary of the observations and data. Section 3 describes the method used to retrieve the transmission spectrum, along with an explanation pf the relevant effects in the RM model. This is followed by an analysis of the white-light fit in the subsequent section. Finally, in section 5, we  examine the transmission spectrum to search for the presence of Rayleigh scattering and heavy metal signatures.

\section{Observations}

Two transits of HD 189733b were observed with ESPRESSO during the nights of August 11 and 31, 2021, as part of the guaranteed time of observation (GTO)  under the program 1104.C-0350(T). ESPRESSO is a high-resolution fiber-fed spectrograph that covers the visible range from roughly $380$ to $788$ nm, distributed over 170 slices (2 slices correspond to one spectral échelle order). They were carried out at UT1 using the HR21 observing mode, with a $1"$ fiber, a spatial binning factor of two, and with a resolving power of approximately $140\,000$.\par
The observations were performed using two fibers, Fiber A pointed to the target, and Fiber B targeted at the sky. The integration time for each observation was set to $300$s during both nights. For the first and second nights, a total of $41$ and $43$ data points were obtained, respectively, resulting in a total effective observation time of 3h 25m and 3h 35m. This resulted in an average SNR (signal-to-noise ratio) of $156$ and $167$, around $580$ nm, and an average uncertainty for the RV measurements of approximately $32$ cm\,s$^{-1}$ on both nights. A summary of the observations is provided in Tab. \ref{tab:observations_summary}, and the plots with the RV measurements, their uncertainties, as well as SNR and airmass variation, can be found in Fig. \ref{fig:header_data}.\par
The data was reduced using the ESPRESSO data reduction pipeline, DRS, version 2.3.1 and 3.0.0. We proceeded with the RVs reduced with 2.3.1 since the latest version is seemingly more prone to jitter. The Cross-Correlation functions were computed using a K2 mask. The RVs were derived by fitting a Gaussian function to the sky-subtracted CCFs for each slice. One spectrum from the end of the second night was removed due to low SNR.
  \begin{table}
\caption{\texttt{CaRM} parameters using the \texttt{SOAP} model.}
\label{table:pars}
\centering
\begin{tabular}{ll}
\hline
Parameter & Physical meaning \\
\hline \noalign{\smallskip}
$V_{sys}\,$(km\,s$^{-1}$) & Systematic velocity of the system \\ \noalign{\smallskip}
$R_{p}/R_{*}$ & Radius ratio between planet and host star \\ \noalign{\smallskip}
$m_P$ (m$_\oplus$) & Planetary mass \\ \noalign{\smallskip}
$a$ (R$_\star$)& Semi-major axis in units of stellar radius \\ \noalign{\smallskip}
$i_p\,$ ($^\circ$)& Orbital inclination \\ \noalign{\smallskip}
$\lambda\,$($^\circ$) & Projected spin-orbit angle \\ \noalign{\smallskip}
$\log(\sigma_{W})$& Logarithm of the jitter amplitude \\ \noalign{\smallskip}
$\Delta \phi_0$ & Mid-transit phase shift \\ \noalign{\smallskip}
\hline
Stellar Properties & \\
\hline \noalign{\smallskip}
$P_{rot}$ (km\,s$^{-1}$) & Rotation period of the host star \\ \noalign{\smallskip}
$V_{CB}$ (km\,s$^{-1}$) & Local convective blueshift amplitude \\ \noalign{\smallskip}
$\alpha_B,\,\alpha_C$ & Differential rotation coefficients \\ \noalign{\smallskip}
$i_\star\,$($^\circ$) & Stellar inclination relative to the sky plane \\ \noalign{\smallskip}
$u_i$ & Limb-darkening coefficients \\ \noalign{\smallskip}
$(T_{eff},\, \sigma_{T_{eff}})\,$ (K)& Effective temperature and uncertainty \\ \noalign{\smallskip}
$(\log(g),\, \sigma_{\log(g)})$& Surface gravity and uncertainty \\ \noalign{\smallskip}
$(z,\, \sigma_{z})\,$ (dex) & Metallicity and uncertainty \\ \noalign{\smallskip}
\hline
\end{tabular}
\tablefoot{The first set corresponds to orbital and planetary properties. The second constitutes the set of stellar parameters, from which $P_{rot}$, $V_{CB}$, $\alpha_B,\,\alpha_C$, $i_\star$ and $u_i$ can be free parameters of the RM modeling. }
\end{table}

\subsection{Simultaneous EulerCam photometry}
We observed two full transits of HD 189733b with the EulerCAM photometer (\citealt{Lendl2012}) at the 1.2m Euler-Swiss telescope located at La Silla observatory. The observations were carried out on 10 August 2021 and 30 August 2021 in the Gunn $r'$ filter with an exposure time of 30~s and 10~s, respectively. The EulerCam data were reduced using the standard procedure of bias subtraction and flat-field correction. The transit lightcurves were obtained using differential aperture photometry, with a careful selection of reference stars and apertures that minimize the final light curve RMS. To account for correlated noise that affects the photometric data due to observational, instrumental and stellar trends, we used a combination of polynomials in several variables (time, stellar FWHM, airmass, coordinate shifts and sky background). The system parameters were obtained using CONAN (\citealt{Lendl2017}, \citeyear{toi222}), a Markov Chain Monte Carlo (MCMC) framework, by fitting for $R_{P}$/$R_{*}$, $b$, $T_{14}$, $P$ and $T_{0}$, assuming wide uniform priors. The quadratic coefficients and their uncertainties for the photometric filter were calculated with the LDCU$\footnote{\url{https://github.com/delinea/LDCU}}$ routine (\citealt{Deline2022}) and allowed them to vary in the fit with Gaussian priors. We also took into account additional white noise by adding a jitter term for each light curve.

\section{Method}
 The development of high-precision spectrographs and sophisticated data has enabled the measurement of RVs with unprecedented precision. Consequently, we can now measure the RM effect with higher detail. However, achieving this level of precision comes with challenges. At the sub-meter per second precision, there are second-order effects in RVs that must be modeled to avoid bias in estimation of orbital parameters.\par
 In this section, we describe how we take advantage of \texttt{CaRM} modularity to retrieve the broad-band transmission spectra. To model the RM effect, we used a version of \texttt{SOAP} \citep[e.g.][]{2012A&A...545A.109B, 2013A&A...549A..35O,2014ApJ...796..132D,2018A&A...609A..21A, 2023A&A...671A..11Z} similar to the described in \citet{10.1093/mnras/staa553}. This is an addition to the already available to use models proposed by \citet{Boue2013} and \citet{Ohta2005} \citep[\texttt{ARoME}  and \texttt{RMcL} implemented by ][]{Czesla2019}. As we will discuss later, this approach allows us to address some approximations made in the previous models that simplify greatly the description of the stellar surface.

\subsection{Modeling the RM effect}
In a rotationally symmetric star without activity, the integrated projected velocity fields towards the observer cancel out exactly the portion associated with the stellar surface rotating away. However, when a planet transits, the planetary disk blocks light from the host star and the stellar surface behind it. Consequently, it becomes possible to measure the integrated velocity imbalance resulting from the unblocked portion of the star. The measured RV amplitude is primarily a function of the area being blocked (planet radius and atmospheric height), the rotation velocity of the star, and the impact parameter \citep{Triaud2018}.\par
Over the years, several attempts have been made to model the classical RM effect with increasing accuracy. Most of these approaches try to express the RM anomaly though analytical formulations, which are constrained by the simplifying assumptions and symmetry conditions. These assumptions are primarily made to make  the integration time to obtain the RV profile becomes practical, but it is also important to note that second degree phenomena create solutions that are not analytically exact. One such limiting assumption is that the underlying "quiet star"\footnote{By "quiet star" we mean a star free of stellar activity and with velocity fields that are generated by its rotational motion.} CCF remains constant across the stellar disk and can only be subject to displacements resulting from the projected rotational velocity (in longitude). As such, no latitudinal variations such as those associated with the differential rotation can be accounted for.\par
To overcome this problem, in this paper, we use an alternative formulation to measure the impact in RVs of a transiting exoplanet using the \texttt{SOAP} code. In short, the code simulates the star as a 2D disk with a grid with a user-defined resolution. We adopt a grid of $600\times600$ which strikes balance between speed and accuracy. For each point on the grid, the code computes the velocity shift to be applied to the "quiet-star" CCF, photometrically scaled to match the limb-darkening at the grid position. This CCF, the default from \texttt{SOAP}, was obtained by cross correlating an observation of the Fourier Transform Spectrograph (FTS) with a G2 HARPS template. The RV measurement results from the Gaussian fit to the sum of all CCFs on the grid. We exploit this numerical point-by-point RV shift computation to include the effect of center-to-limb variations (CLVs) induced by the convective blueshift (CB) and differential rotation. This changes and additional updates will be presented on a forthcoming publication of \texttt{SOAP}.

 \subsection{Convective Blueshift}
 The convective blueshift \citep{1896ApJ.....3...89J} is an RV signal that arises from the granular nature of stars with an external convective layer. On average, the hot and bright rising plasma in the granules contributes more significantly to the integrated radial velocities compared to the darker and cooler gas that sinks in the inter-granular spaces. The first degree changes induced by the radial component of the CB are the result from the projection effect along the disk and the photometric weight of the limb darkening.
 Similar to the rotational velocity fields, a transiting exoplanet introduces an unbalance in the perceived projected velocities, due to the CB, which is superimposed on the RM effect. \par
 To incorporate the effects of the CLVs induced by the CB, we implemented in \texttt{SOAP} the first-order approximation of the effect described by \citet{2011ApJ...733...30S}. We note, however, that more sophisticated approaches exist and can potentially better model the CB effect. However, these approaches involve additional physics (and the corresponding assumptions) based on magnetohydrodynamical simulations of the stellar surfaces \citep{2016ApJ...819...67C}. We consider that the first-degree polynomial description of the CLVs is a good compromise between model complexity and capturing the bulk of the CB effect. Nonetheless, this model has significant limitations, as it neglects the presence of meridional flows, variations in the CB strength and line shape along the disk for different chemical species \citep[e.g.][]{2021A&A...654A.168L} and differential rotation. We expect this approximation to be reasonable for slow rotating solar-type stars, where the FWHM is greater than the rotation speed. Despite, even within this group, significant deviations from the profiles given by MHD models are possible \citep[e.g.][]{2016ApJ...819...67C}.\par
 To compute the CB-induced CCF shifts, we consider an initially limb-darkened perfect sphere. In each cell SOAP grid, we assume a constant CB velocity perpendicular to the surface. The magnitude at each point is radially symmetric and results from the projected component by the angle between the normal to the stellar surface and the line of sight (LOS) $\theta$: $V_{CB} \cos{\theta}$ where $V_{CB}$ is the local CB velocity. In the literature, the measurement of the convective blueshift velocity is often represented by the solar-scale factor $S$ \citep[e.g.][]{2021A&A...654A.168L}. This quantity represents the ratio between the CB amplitude of the star and that of the Sun, where the solar value corresponds to $-350$ m\,s$^{-1}$.

 \subsection{Differential rotation}
 Differential rotation was initially detected on the Sun through the observation of variations  in the migration rate of spots at different latitudes.Spectroscopically, it manifests as a latitude-dependent shift of the spectral lines, resulting in measurable RV shifts \citep[e.g.][]{1969SoPh....9..448L}. \par
 Solar-like star, are known to possess strong magnetic fields that give rise to activity that is manifested on the surface \citep{1908ApJ....28..315H}.In alpha-omega dynamos, such as the Sun, magnetic fields are generated by the dynamo that is located within the convective envelope \citep{1955ApJ...122..293P}. The dynamo mechanism itself relies on the presence of turbulent velocities and differential rotation, both of which are observed in the Sun.\par
To account for the latitudinal effect of the differential rotation, we incorporate the formulation presented in, for example, \citet{2003A&A...408..707R} or \citet{gray_2005}:
 \begin{equation}\label{eq:veq}
 \Omega(l) = \Omega_{eq}(1-\alpha_B \sin^2(l)),
 \end{equation}
 where $\Omega(l)$ represents the angular velocity component at the latitude $l$, $\Omega_{eq}$ denotes the angular rotation velocity at the equator, and $\alpha_B$ represents the differential rotation coefficient. For a given star with differential rotation, the impact on the RM profile is not only greater in amplitude for higher coefficients but also more pronounced (and less degenerate with the other parameters) if the planet's transit path traverses a wider range of stellar latitudes \citep{2022A&A...661A..97R}.

 \begin{table}
\centering
\caption{Set of priors for the white-light fit with \texttt{CaRM} for HD 189733b.}
\label{table:white_priors}
\begin{tabular}{lll}
\hline
\hline
Parameter & Prior \\
\hline
$V_{sys}\,$\tablefootmark{$\dag$} (km\,s$^{-1}$) & $\mathcal{G}(-2.18, 0.1)$\\
$R_p /R_\star $ & $\mathcal{U}(0.14, 0.17)$\\
$\Delta \phi_0$ \tablefootmark{$\dag$} & $\mathcal{G}(0, 0.005)$\\
$m_p$ \tablefootmark{$\dag$} ($m_\oplus$) & $\mathcal{G}(363, 10)$\\
$\alpha_B$ & $\mathcal{U}(0, 1)$\\
$i_\star$ ($^\circ$)& $\mathcal{U}(0, 180)$\\
$i_p$ ($^\circ$)& $\mathcal{G}(85.5, 0.1)$\\
$V_{CB}$ (m\,s$^{-1}$)& $\mathcal{U}(-350, 0)$\\
$\lambda$ ($^\circ$)& $\mathcal{G}(-0.85, 0.32)$\\
$P_{rot}$ (days)& $\mathcal{G}(11.953, 0.1)$\\
 & $\mathcal{U}(7, 13)$ \\
$a$ ($R_\star$)& $\mathcal{G}(8.756, 0.0092)$\\
$\sigma_{W} $\tablefootmark{$\dag$} \tablefootmark{*} (m\,s$^{-1}$)& $\mathcal{U}(10^{-4}, 1) $\\
\hline
\end{tabular}
\tablefoot{
In the table, the symbol $\mathcal{G}$ represents a normal distribution, where the first value corresponds to the mean and the second the standard deviation. Similarly, the symbol $\mathcal{U}$ represents a uniform distribution, with the respective lower and upper boundaries.\\ \tablefoottext{*}{We present the prior for $\sigma_{W}$ in units of m\,s$^{-1}$ instead of the logarithmic, as it is more intuitive to perceive the range in velocity units.}\tablefoottext{$\dag$}{The fitting process is performed independently for each dataset.}}
\end{table}
 \subsection{\texttt{CaRM}}
\texttt{CaRM}\footnote{https://github.com/EduardoCristo/CaRM} is a semi-automatic code written in \texttt{Python}, designed to  extract the broad-band transmission spectra of exoplanets using the chromatic RM method \citep{2022A&A...660A..52C}. The code takes as input CCF files from HARPS and ESPRESSO (which are products from the default DRS) or user-specified text files containing the order/slice ranges, the observation times, radial velocities, and their uncertainties (organized by columns). \par
To initialize the code, the user modifies a setup file, which includes the path to the folders containing the reduced data (the CCF files) that correspond to the observations. \texttt{CaRM} scans the folders for CCF files with a user-defined suffix. It then converts the data to the specifies text file structure, employing lists with the spectral format of each instrument. At this step, the RVs are computed by performing a Gaussian fit to the weighted sum, by the variance, of the CCFs as defined in the range list. The uncertainties are computed assuming a photon-noise-limited observations, following the method described in \citet{Bouchy2001} but adapted to directly measure them from the CCF.\par
The input file allows the user to choose which model use to perform RM RV anomaly fit. Stellar parameters such as effective temperature, stellar surface gravity, and stellar metallicity are not directly  fed into the models but are used instead to fit the limb-darkening coefficients using\texttt{LDtK} \citep{Parviainen2015}.\par
The code performs the model fitting using wither MCMC implementation \texttt{emcee} \citep{ForemanMackey2013} or a dynamic nested sampler \texttt{Dynesty} \citep[e.g.][]{2020MNRAS.493.3132S, sergey_koposov_2022_7148446}. Priors are defined as a dictionary, associating each parameter to be fitted with its prior distribution. Joint fits of specific parameters can be performed if multiple data sets are provided. The same prior definition scheme is employed for the subsequent fits performed on different wavelength ranges. The code assumes that the first element of the range list corresponds to the white light, and it uses this data to compute the color-independent parameters with the highest SNR. The transmission spectrum is then constructed, after each chromatic RM variation is performed, with the measurements of the planet-to-star ratio as a function of the wavelength.
\begin{figure}
    \centering
    \includegraphics[width=\linewidth]{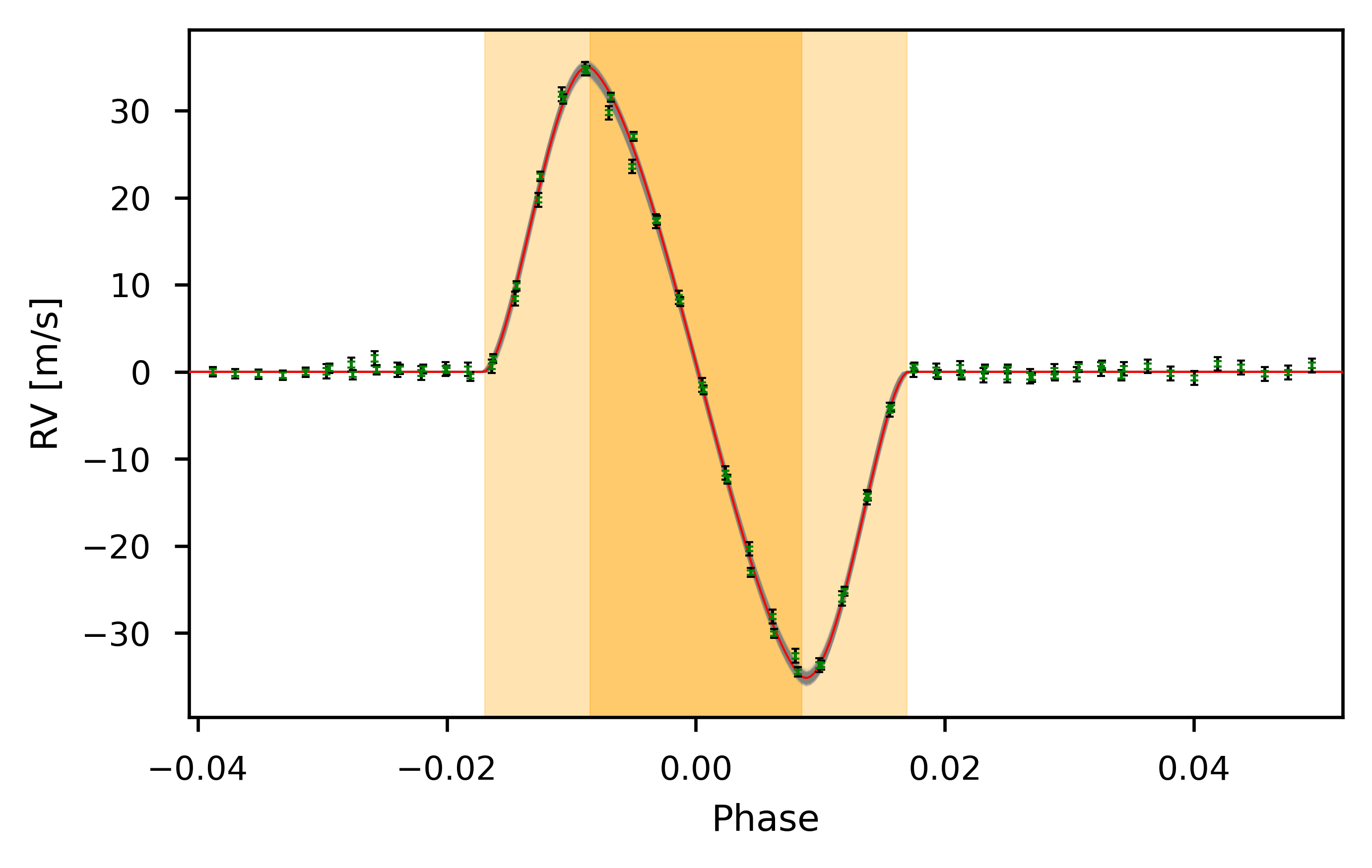}
    \includegraphics[width=\linewidth]{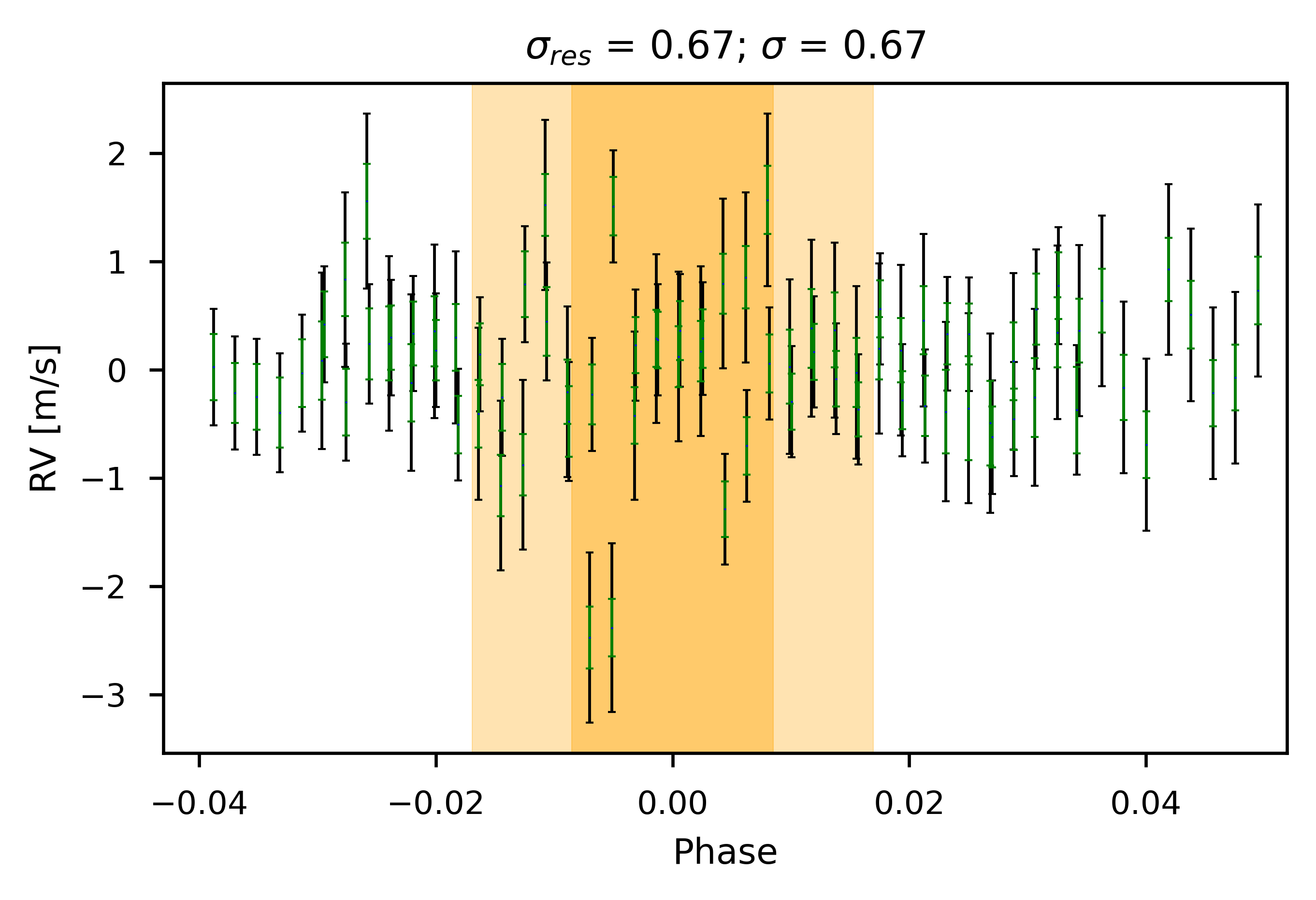}
    \caption{Fit and residuals of the white light data. Top: The best-fit model (solid red line) obtained from the combined data of the nights observed with ESPRESSO. Bottom: Residuals after subtracting the model. The light orange areas represent the phases where the planet is fully inside the stellar disk, while the lighter orange regions correspond to the ingress and egress phases. The data points are represented with two error bars. The green error bar is computed from the CCF, assuming a photon-noise limited observation. The black error bars is obtained by adding in quadrature the value of the green error bars and the jitter amplitude. At the top of the bottom figure, two quantities are presented. The first is the average dispersion of the residual RVs ($\sigma_{res}$) in units of $m\,s^{-1}$. The second is the average value of the black error bars, also given in $m\,s^{-1}$.}
    \label{fig:hd189jres}
\end{figure}

\section{The white light fit}
The chromatic Rossiter-McLaughlin effect relies on accurately fitting of the RM anomaly as a function of wavelength. Therefore, we begin by fitting the white light RVs (which  correspond to the full bandpass of ESPRESSO) to constrain the color-independent parameters (Tab. \ref{table:pars}) with the highest SNR the data can offer.\par

\begin{figure*}[ht!]
    \centering
    \includegraphics[width=0.49\linewidth]{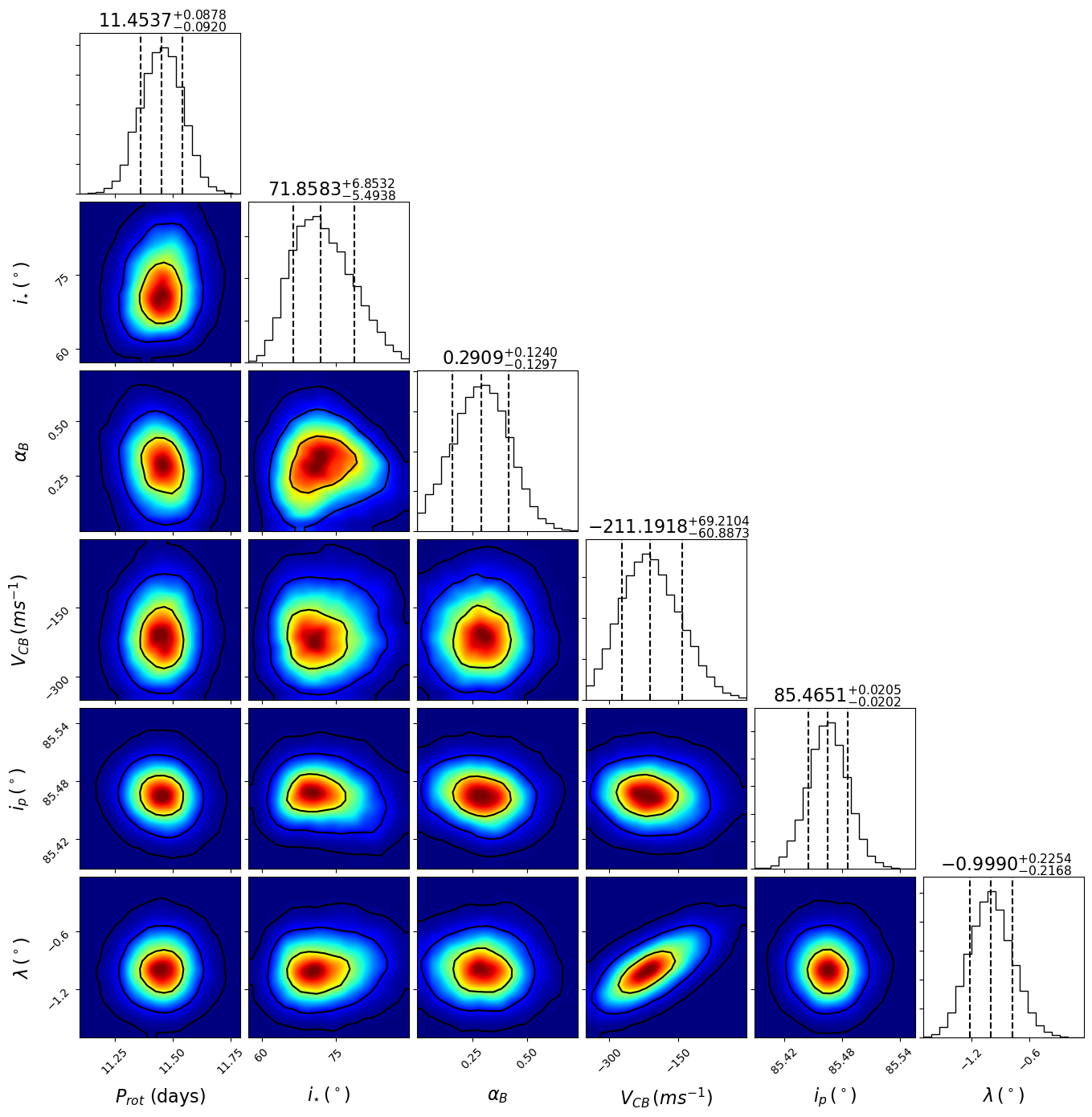}
    \includegraphics[width=0.49\linewidth]{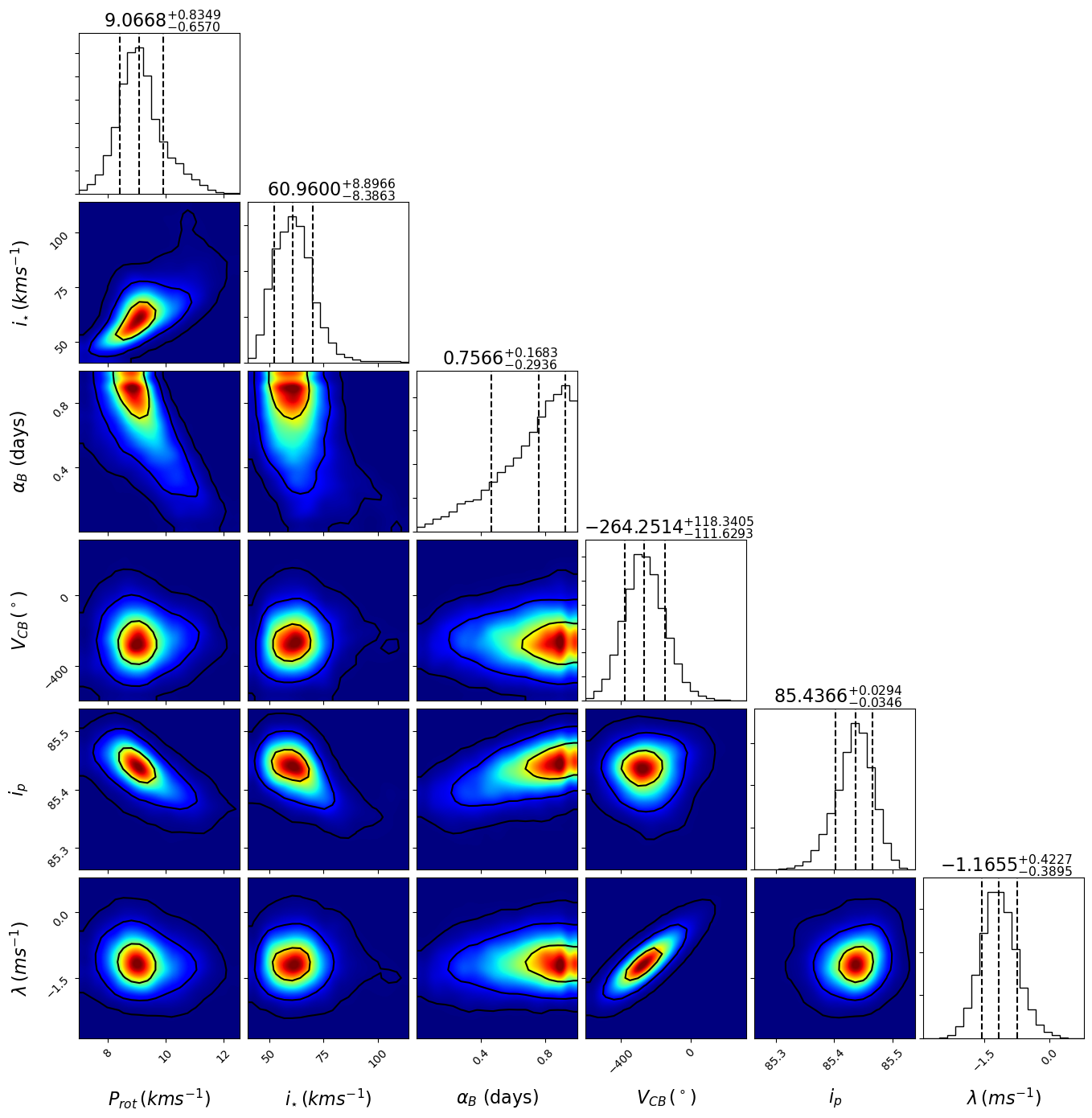}
    \caption{Posterior distribution diagrams for M1 and M2. The corner plots depict the posterior distribution for the equatorial rotation period, stellar axis inclination, differential rotation coefficient, local CB velocity and projected spin-orbit angle. The black contour lines represent (from center to outwards) the confidence intervals enclosing $68.27\% \, 95.45\%$ and $99.73\%$ of the accepted samples. The histograms display the parameter posterior distributions, with the darker dashed line indicating the median value lighter lines delimiting the 1-sigma interval.}
    \label{fig:corner_difrot_clvs}
\end{figure*}
\subsection{Priors and assumptions}
For the white-light fit, we assumed a circular orbit that is expressed by a Keplerian with semi-amplitude $K$ and a systemic velocity $V_{sys}$.We allow $V_{\text{sys}}$ to vary as a free parameter with a uniform prior, as it can be influenced by nightly offsets, telluric contamination, and stellar activity, which may alter its measured value\footnote{There is extensive discussion in the literature about HD 189733 being an active star, in particular, chromospheric activity \citep[e.g.][]{Bouchy2005, Boisse2009}.}.
The out-of-transit slope, associated with the Keplerian orbit of the planet, and often parameterized by its  semi-amplitude K, is known to be affected by stellar activity \citep{Boldt2020}. Notably, $K$ is not given directly in \texttt{SOAP} but as a function of the planetary mass. Given that we expect the planetary mass to remain constant during the transit, the Gaussian prior only accommodates the impact of stellar activity.
 For the mid-transit time, a shift term is introduced with a Gaussian prior. The mean value of the prior was set to zero, and a standard deviation was chosen to be compatible with the timescale of the transit. The mid-transit shifts can be originated from the uncertainties in the determination of the mid-transit time, which are amplified by the number of orbits since the epoch used as reference. Although we anticipate this value to be minimal due to extensive studies of this planet, other sources of mid-transit shifts, such as small-scale transit time variations (TTVs) or small orbital eccentricities, cannot be ruled out. We adopt uniform priors for $\alpha_B$, $i_\star$, $V_{CB}$, and $\sigma_{W}$ since these parameters are poorly constrained in the literature or unknown for this target or these particular observations. Gaussian priors based on literature values were used for the remaining parameters, with the standard deviation matching the reported uncertainties. The width of the $P_{rot}$ prior was increased since it is used to compute the rotational velocity of the star, which presents some variability in the literature (model 1 or M1). It is important to note that in our analysis, we selected a source from the literature that obtained the rotation period value using an alternative method based on the evolution of stellar spots. This choice was made because it is possible that fitting the RM models alone may not be sufficient to fully resolve the degeneracies between the parameters in this well-aligned planetary system. By incorporating additional constraints from spot evolution studies, we aim to improve the accuracy and reliability of our results. In addition, we also tested a broader prior for the rotation period to access rotation periods that significantly deviate from the value that we used as reference before (model 2 or M2). Table \ref{table:white_priors} summarizes and completes the description of the set of free parameters and the respective priors we adopt.\par

\subsection{Results}

We run \texttt{CaRM} selecting \texttt{Dynesty} to perform the fit using a nested sampling approach. The number of live points is set to $500$. The convergence is evaluated by monitoring the estimate of the logarithmic evidence and terminating the run when the variation is below $1\%$. This yielded $33\,056$ posterior samples for model M1 and $38\,336$ for model M2 during the static sampling phase. The corner plots illustrating  the posterior distributions of the fitted parameters can be found in Fig. \ref{fig:corner_white} and Fig. \ref{fig:corner_white_m2}. The fits of the individual nights with the best-fit model and corresponding residuals after subtracting it are in Fig. \ref{fig:hd189_individual}. The fit for M2 is similar to M1 but they show higher dispersion ($80$ cm\,s$^{-1}$). Based on this, we decide to adopt M1 for further study. It is important to note, however, that achieving lower residuals at the expense of over-fitting is a potential concern. To address this, we have verified that the dispersion of the residuals for M1 is still more than twice the expected photon noise.\par
There is no significant observable correlation between the residuals of the first and second nights. It exists, however, points that deviate significantly from the average RVs, which could be due to occultation events of active regions on the stellar surface, such as spots or plages.Notably, a larger radial velocity variation can be observed on the second night just after ingress, but there is no corresponding variation in the first night. The signal of the RV variation suggests a possible spot-crossing event. \par
To investigate the origin of the significant deviation in the residuals of the first night, we utilized simultaneous ESPRESSO and EulerCam observations. We found a good agreement both in magnitude and phase, b by modeling the residual radial velocities using \texttt{SOAP} with an occultation of a single spot at a stellar longitude of $-30^\circ$ and latitude of $0^\circ$. We assumed spot size of $0.1 \%$ and a temperature contrast of $660K$. This produces a negative RV variation with  an amplitude of approximately $3$ m\,s$^{-1}$ and a corresponding positive flux excess of $1300$ ppm. For the photometric measurements, we rebinned the flux observations to approximately match the RVs at the same phase, as shown in Figure \ref{fig:hd189jres}. The photometric measurements are compatible with the spot occultation scenario. However, since the effect in the flux is low when compared with the average error bars, we cannot rule out other possibilities.\par
\begin{figure*}
\centering
\includegraphics[width=0.497\linewidth]{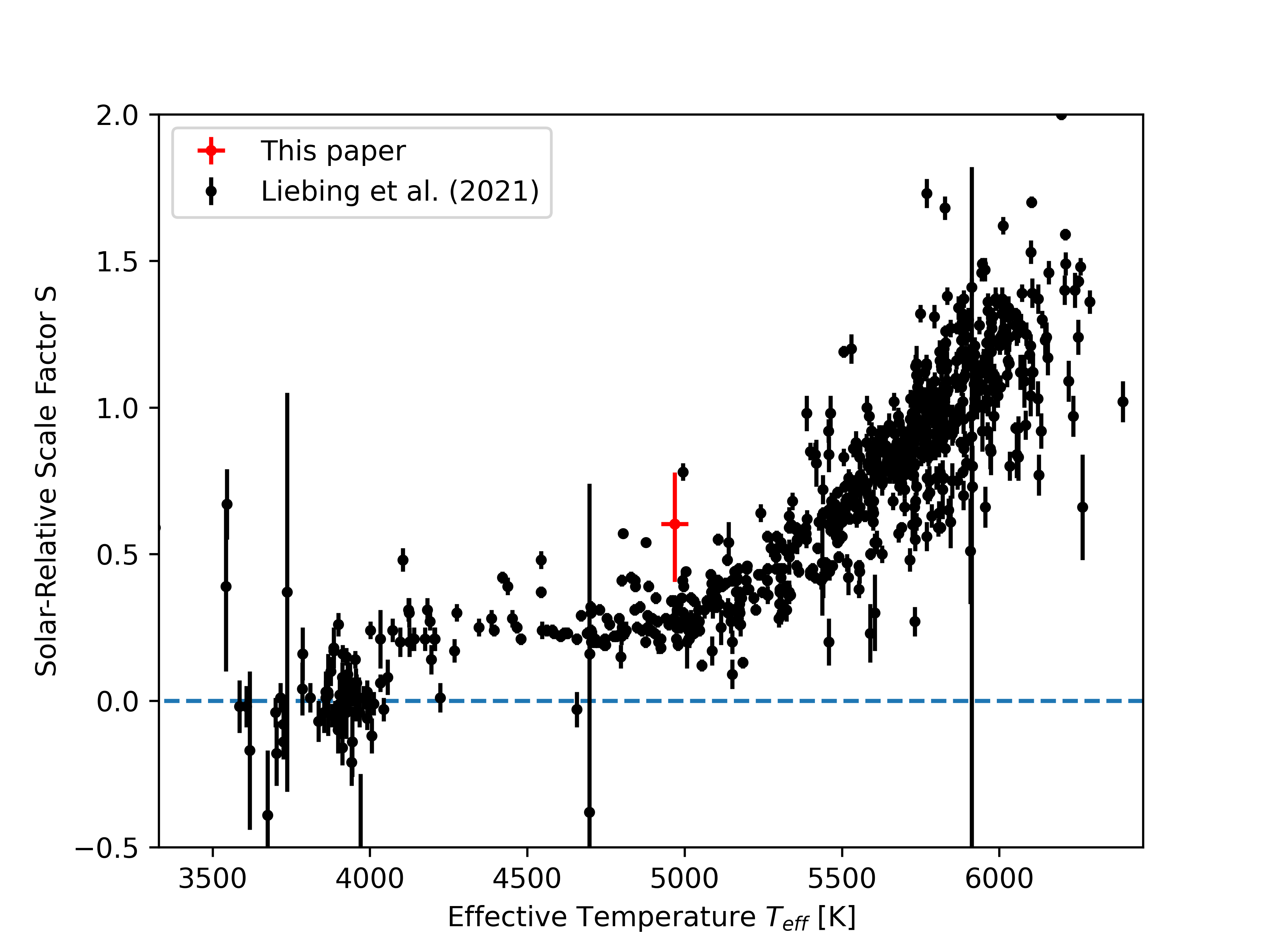}
\includegraphics[width=0.497\linewidth]{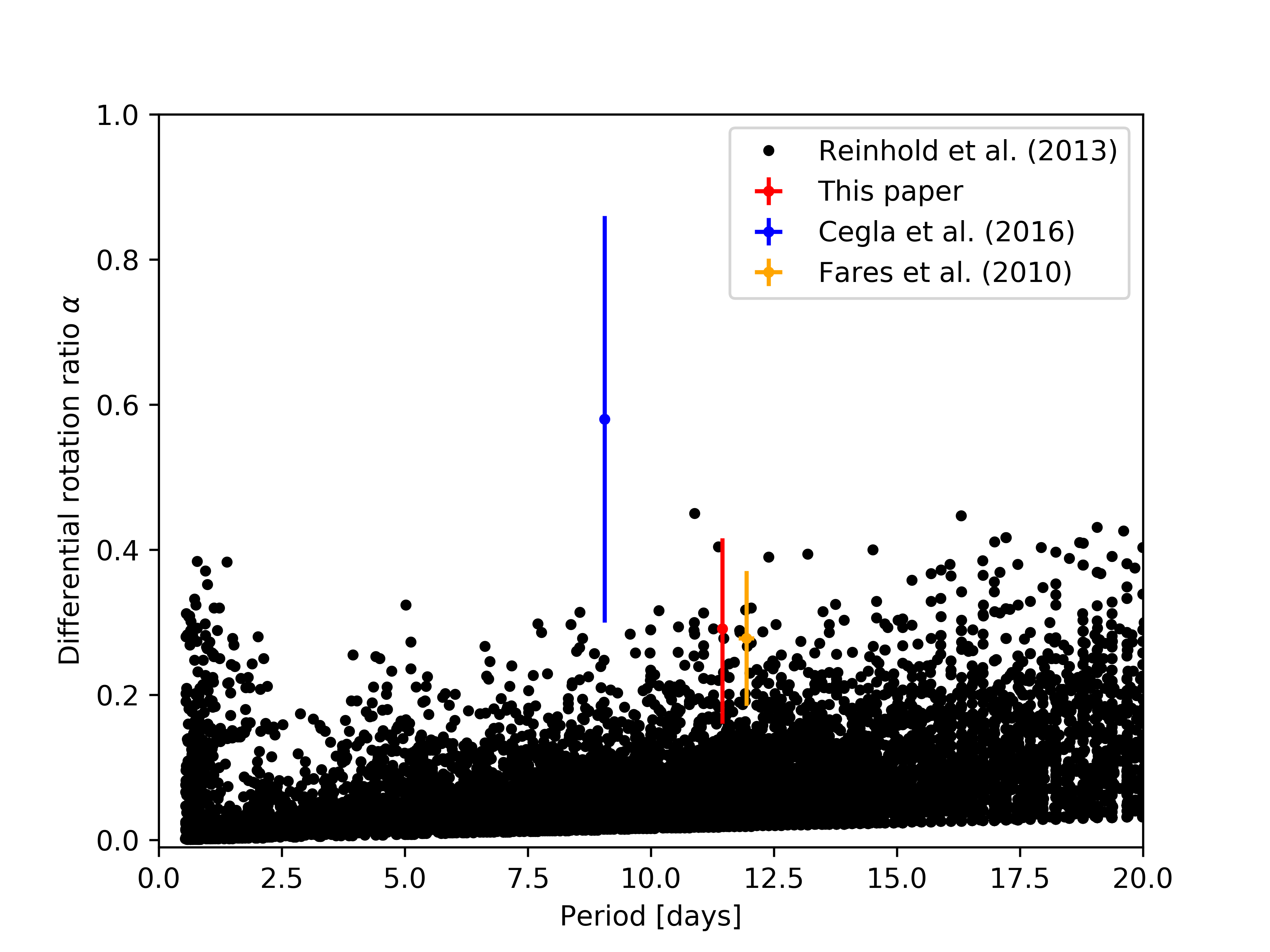}
\caption{Estimates of the solar-relative CB and differential rotation ratio, along with a comparison to existing literature. Left: The solar-relative scale factor as a function of the effective temperature for a sample of stars ranging from M to F spectral types \citep{2021A&A...654A.168L}. The red point and error bars represent the prediction from our best-fit model. Right: Relative differential rotation coefficient as a function of the rotation period of the stellar surface at the equator. We compare our retrieval with the results from \citet{2010MNRAS.406..409F} and \citet{Cegla2016}. Additionally, we compare these measurements with the differential rotation coefficients obtained from the analysis of photometric data of 24,124 Kepler stars \citep{2015A&A...583A..65R}. We represent these measurements as black dots. For the period, we selected the minimum one provided by the authors since it corresponds to the equatorial period for stars exhibiting solar-like differential rotation.}
\label{fig:CLV_diffrot}
\end{figure*}

Figure \ref{fig:hd189jres} displays the best-fit model obtained from combining the data of the two nights, as well as the best fit to the observed data. The joint model fit of the nights results in a RV scatter of $67$ cm\,s$^{-1}$, which is approximately twice the predicted photon noise level. The error bar that includes the jitter in quadrature closely matches this value.  By comparing our retrieval with these previous studies, we demonstrate the significance of incorporating models that account for center-to-limb variations (CLVs) and differential rotation to avoid potential biases \citep{2016ApJ...819...67C, Cegla2016}. Additionally, the presence of strongly correlated signals in the residuals has been highlighted in previous works \citep[e.g.,][]{Triaud2009, Cegla2016, 2022A&A...660A..52C}.

\begin{table}
\centering
\caption{Posterior distributions of the white-light fit with \texttt{SOAP} for M1}
\label{table:white_posterior}
\begin{tabular}{lll}
\hline
\hline
Parameter & Posterior \\
\hline
$V_{sys, \,  0} \,$ (km\,s$^{-1}$) & $-2.18095_{-0.00011}^{+0.00010}$\\
$V_{sys, \, 1} \,$ (km\,s$^{-1}$) & $-2.18675_{-0.00014}^{+0.00014}$\\
$P_{rot}$ (days)&  $11.454_{-0.088}^{+0.092}$\\
$i_\star$ ($^\circ$)& $71.87_{-6.91}^{+5.55}$\\
$\alpha_B$ & $0.29_{-0.12}^{+0.13}$\\
$V_{CB}$ (m\,s$^{-1}$) & $-211_{-69}^{+61}$\\
$i_P$ ($^\circ$)& $85.465_{-0.021}^{+0.020}$\\
$\lambda$ ($^\circ$)& $-1.00_{-0.23}^{+0.22}$\\
$a$ ($R_\star$)& $8.7686_{-0.0082}^{+0.0082}$\\
$R_p /R_\star $ & $0.1602_{-0.0035}^{+0.0039}$\\
$m_{p, \,  0} \,$ ($m_\oplus$)& $356.9_{-1.2}^{+1.2}$\\
$m_{p, \,  1} \,$ ($m_\oplus$)& $352.1_{-1.4}^{+1.5}$\\
$\Delta \phi_{p, \,  0} \,$ & $-0.002424_{-0.00035}^{+0.00036}$\\
$\Delta \phi_{p, \,  1} \,$ & $-0.002300_{-0.00044}^{+0.00044}$\\
$\sigma_{W, \,  0} \,$ (cm\,s$^{-1}$)& $50.4_{-9.1}^{+8.9}$\\
$\sigma_{W, \,  1} \,$ (cm\,s$^{-1}$)& $72.5_{-10.1}^{+12.6}$\\

\hline
\end{tabular}
\tablefoot{The uncertainties represent the $68.27\%$ confidence interval around the median value. \tablefoottext{*}{We provide here the prior for $\sigma_{W}$, in m\,s$^{-1}$, and not the logarithmic since it is more intuitive to perceive the range in velocity units}\tablefoottext{$\dag$}{Independently fit for each data set.}}
\end{table}

\subsection{The stellar surface and true spin-orbit angle}

The stellar surface has a significant impact on the shape of the RM curve, as we discussed before. Therefore, it is essential to analyze how effectively we can retrieve the fundamental parameters that describe using RV observations modeled with \texttt{SOAP}.\par
Figure \ref{fig:corner_difrot_clvs} illustrates the posterior distribution diagrams for several keys parameters that are used to model the stellar surface. In particular, the equatorial rotation period of the star, the stellar rotation axis inclination, the differential velocity ratio, the local convective blueshift amplitude velocity, and the spin-orbit angle for M1 (right plot).  \par
Our analysis yields a rotation period $P_{rot}\approx 11.45 \pm0.09$ days. Assuming a spherical star, this can be translated converted in linear rotation velocity dividing the equatorial perimeter by the rotation period. We estimate  $V_{eq}= 3.38 \pm 0.06$ km\,s$^{-1}$. Since HD 189733b's orbit is aligned the planet crosses only a small number of stellar latitudes, which can be approximated reasonably well by the average latitude. To compare our results with other studies, we compute the average velocities of the stellar surface behind the planet, using the planetary and stellar inclinations, as well as the differential rotation of the stellar surface at the average latitude crossed by the planet. We approximate the latitudes crossed by the planet using the impact parameter, following a similar approach to \citet{Cegla2016}, with the expression $V_{eq} \sin i_{\star} (1-\alpha (a \cos i_P)^2)$. \par
Measurements of the projected rotation velocity for HD 189733 vary significantly in the literature. For instance, stellar activity photometric modelling yields $2.97 \pm 0.22$ km\,s$^{-1}$ \citep{Winn2006}, while line broadening analysis by \citet{Bouchy2005} suggests $3.5 \pm 1$ km\,s$^{-1}$. 
Our approach is similar to the one adopted by \citet{Triaud2009}\footnote{The author uses, additionally, photometric data to perform a simultaneous fit with RV measurements (including the RM). } where $V_{eq}\sin(i_\star)$ is computed from an RM model. In Triaud's paper, however, their best solution for the rotational velocity ($3.316^{+0.017}_{-0.068}$ km\,s$^{-1}$) produces clear wave-like residuals and they don't consider that the star may have differential rotation or CB. The authors adjust the RM model to account for the observed trends, which results in a lower estimate $V_{eq}\sin(i_\star) = 3.05$. Also, \citet{Cegla2016} applied an analytic model of the Doppler shifts behind the planet, which includes the effect of CLVs and differential rotation, and fitted it to the residual CCFs. They find $V_{eq}= 4.50_{-0.49}^{+0.51}$ km\,s$^{-1}$ and $V_{eq}\sin(i_\star) \approx 3.3$ km\,s$^{-1}$. Our solution for the sky-projected spin-orbit angle ($\lambda$) of HD 189733b is consistent with previous literature. We estimate $\lambda = -1.00_{-0.23}^{+0.22}$, which is in agreement with values ranging from $-1.4\pm1.1^\circ$ reported by \citet{Winn2006} to $-0.35\pm0.25^\circ$ reported by \citet{2016ApJ...819...85C}. These results suggest a higher statistical probability for the stellar axis to be close to the orbital inclination. Statistical analyses of spin-orbit misalignment, such as those conducted by \citet{2016ApJ...819...85C} or \citet{2009ApJ...696.1230F} provide valuable insights into the distribution of spin-orbit angles and their implications. Our finding of $\lambda \approx -1.00$ supports the notion that the stellar axis is more likely to align closely with the orbital inclination, based on these statistical analyses.\par

In our model, we fit the stellar rotation axis angle. This is an important measurement to constrain models of planetary formation and evolution. For example, the angle between the Sun’s rotation axis and the ecliptic plane is  $7.15^\circ$ \citep[e.g.][]{2005ApJ...621L.153B} which represents only a small deviation from it. For the HD 189733 system, we retrieve a stellar axial rotation tilt of $71.87^{ +6.91^\circ}_{-5.55^\circ}$. This result is in line with the prediction from \citet{Henry_2007} which computes $i_\star>54^\circ$ with $95\%$ confidence, with a most probable value of $65^\circ$. An additional source for measurements of the stellar inclination axis can be found in \citet{Cegla2016}. Comparing the results for the model similar to ours (one parameter CB and differential rotation), they find $92.0^{+11.0^\circ}_{-3.8^\circ}$ which is significantly different from our prediction as their posterior distribution clearly prefers higher values for the angle. \par
We additionally derive the differential velocity ratio $\alpha_B \approx 0.29^{+0.12}_{-0.13}$ with $93.4\%$ confidence for $\alpha_B> 0.05$. This means that, at the poles, the star rotates with a velocity $2.58 \pm 0.35$ km\,s$^{-1}$ which corresponds to a rotational period of $14.9 \pm 2.0$ days. In \citet{Cegla2016} we can find both one and two parameters (using a similar parametrization as can be found on Eq. \ref{eq:veq}) models to describe the latitudinal change of the stellar surface velocity. The authors find $\alpha_B>0.1$ with $99.1\%$ confidence and amplitude that ranges between $0.3$ and $0.86$ with one parameter (derived from HARPS data). A more precise determination is available on \citet{2010MNRAS.406..409F}, where it is derived  $\alpha_B \approx 0.278 \pm 0.093$ from polarimetry. We compare these measurements with our solution and the observed distribution of differential rotation with the stellar rotational period in Fig. \ref{fig:CLV_diffrot} and find a good agreement with the literature values.\par

HD 189733 is a K dwarf and, as such, it is expected to have a granular surface like the Sun but with a lower contrast between the granule centers and the inter-granular lanes. Using the formulation for the CLVs produced by the CB, the local convective blueshift must be sub-solar (or $S$ factor lower than one) \citep{2021A&A...654A.168L}. We are able to measure $V_{CB} \approx -211^{+69}_{-61}$ m\,s$^{-1}$ which represents a scale factor $S \approx 0.60^{+0.20}_{-0.18}$ which is sub-solar as expected. Despite the value, in RV amplitude for the CLVs created by the CB, being lower than the differential rotation, it contributes to a deformation symmetrical about the mid-transit to the RM profile (Fig. \ref{fig:dopplermap}). This characteristic allows a confident determination of the value for the CB itself, with $V_{CB}<-50$ m\,s$^{-1}$ with a level of $99.8\%$. We compare our result with \citet{2021A&A...654A.168L} in Fig. \ref{fig:CLV_diffrot}. We depict the Doppler maps generated by the combination of the effects described in this section, as well the contribution of each of them, in Fig. \ref{fig:dopplermap}.\par

\section{Transmission spectrum}

\begin{figure*}[ht!]
    \centering
    \includegraphics[width=\linewidth]{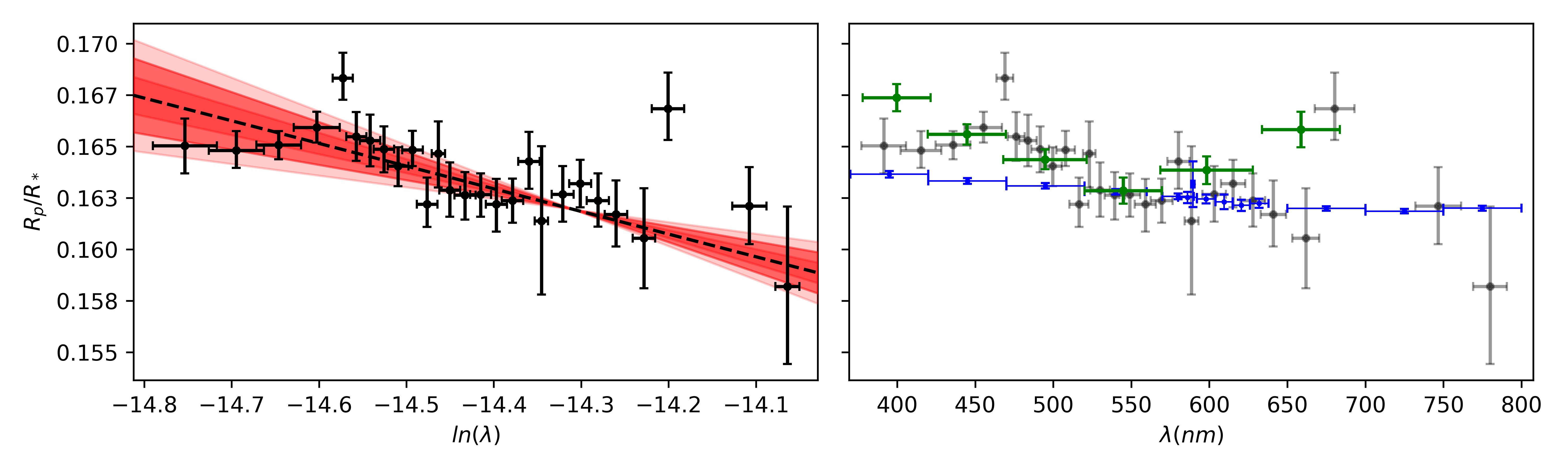}
    \caption{Broad-band transmission spectrum of HD 189733b. Left: Linear fit (black dashed) to the measured planet-to-star radius ratio as a function of the logarithmic wavelength. The red bands represent the one to three $\sigma$ confidence levels. Right: Comparison of our results with \citet{Pont2013} in blue and \citet{2022A&A...660A..52C} in green. Shifts were applied to the literature results to match, approximately, the radius level @550nm. }
    \label{fig:transmission_compare}
\end{figure*}

\subsection{Chromatic RM fit}
Similarly to the white light, we computed the RVs that results from the sum of CCFs that are defined in the specific slice intervals (see Tab. \ref{tab:radius_table}). These RVs reflect, in practice, the impact of the planetary transit on the stellar spectrum, which results in an RM profile that captures the sum of the planetary radius and scale height for the particular wavelength interval. \texttt{CaRM} fits these chromatic Rm\,s using a priori information from the white light. We fix the wavelength-independent parameters such as the differential rotation velocity ratio, rotational period, or planetary inclination such that biases are not introduced in the chromatic radius determination (the RVs here have a substantially lower SNR). We fit similarly the systemic velocity of the star for each night with a uniform prior to account for a potential wavelength-dependent offset due to stellar activity. For the same reason stated for the white light, in addition to the chromatic variations, we let the planetary mass change around the literature value with comprehensive $10\%$ width. We also give a uniform prior to the convective blueshift since it is expected to be a function of the wavelength \citep{Cegla2016}, as it results from the contribution of different spectral lines that are formed at different depths in the photosphere (and, as such, with different velocity contributions). To the jitter amplitude, it is attributed a broader prior, when compared with the white-light, since each bin contains less RV information. For the RM fit bin-by-bin approach used here, we assume that the stellar spectrum shape doesn't change as a function of the distance to the disk center \citep[e.g.][]{2021A&A...649A..17D}. There may be additional parameters that change as a function of the wavelength that we are not considering. The set of priors for the chromatic priors are summarized in Tab. \ref{table:chromatic_priors} and the fit for each wavelength bin is represented in Fig. \ref{fig:chromatic_rm_fits}.\par

\begin{table}[h!]
\centering
\caption{Set of priors for the chromatic fit. The elements of the table as the same meaning as Tab. \ref{table:white_priors}.}
\label{table:chromatic_priors}
\begin{tabular}{lll}
\hline
\hline
Parameter & Prior \\
\hline
$V_{sys}\,$\tablefootmark{$\dag$} (km\,s$^{-1}$) & $\mathcal{U}(-2.5, -1.5)$\\
$R_p /R_\star $ & $\mathcal{U}(0.14, 0.17)$\\
$m_P$ \tablefootmark{$\dag$} ($m_\oplus$)& $\mathcal{G}(363, 36)$\\
$V_{CB}$ (m\,s$^{-1}$)& $\mathcal{U}(-1500, 100)$\\
$\sigma_{W}$\tablefootmark{$\dag$} (m\,s$^{-1}$)& $\mathcal{U}(10^{-4}, 10) $\\
\hline
\end{tabular}
\tablefoot{The elements of the table as the same meaning as Tab. \ref{table:white_priors}. \tablefoottext{$\dag$}{Independently fit for each data set.}}
\end{table}

\subsection{Transmission spectrum retrieval}
The transmission spectrum derived from ESPRESSO data, Fig. \ref{fig:transmission_compare}, shows a steep decrease in planetary radii as a function of increasing wavelength. In the wavelength range where they coincide, the transmission spectrum derived in this paper is globally consistent with the spectrum obtained from HARPS data with the same technique \citep{2022A&A...660A..52C}. Our retrieval, however, suggests an enhanced slope when compared with the spectrophotometric spectrum obtained with the observations from Hubble (STIS).\par
The observed slope is classically attributed to the scattering of particles with a size smaller than the incoming light's wavelength. In literature \citep[e.g.][]{des2008rayleigh}, this radius-wavelength slope is often parametrized as:

\begin{equation}\label{eq:eq1}
    \frac{dR_p}{dln(\lambda)}=\alpha H,
\end{equation}    
where $\alpha$ corresponds to the scattering slope and $H$ to the atmospheric scale-height. When $\alpha=-4$ this kind of interaction is often called as Rayleigh scattering. There are several exoplanets that exhibit 'Super-Rayleigh' (or $\alpha$<-4) slopes, such as the hot Jupiter WASP-19b with $\alpha \sim -35$ \citep{2017Natur.549..238S} or the super-Neptune HATS-8b with  $\alpha \simeq -26 \pm 5 $ \citep{May_2020}. Studies of the planet population of \citet{Sing2015} show that, in general, planets have an enhanced slope \citep[e.g.][]{2019MNRAS.482.1485P,2019ApJ...887L..20W} characterized by $\alpha \lesssim -5$. The mechanisms proposed to produce these increased slopes are varied and range from both occulted \citep[e.g.][]{Oshagh2014, Boldt2020} and/or unocculted stellar activity \citep[e.g.][]{2014ApJ...791...55M, Kasper2018ATS,2019AJ....157...96R}, small dimension condensates \citep[e.g.]{2019MNRAS.482.1485P} or photochemical haze \citep[e.g.][]{2019ApJ...877..109K,2020ApJ...895L..47O}.\par
\begin{figure*}
    \centering
    \includegraphics[width=\linewidth]{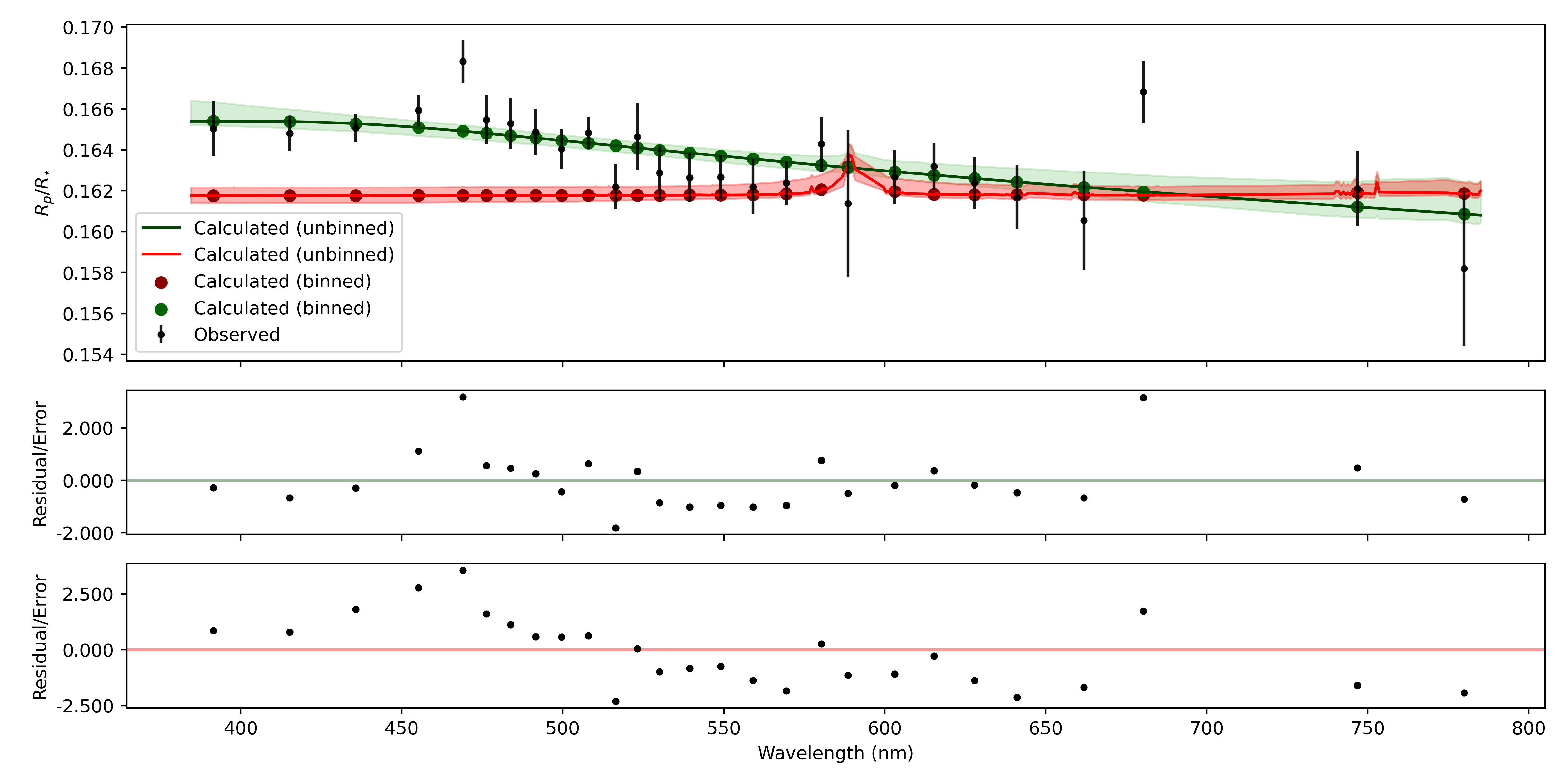}
    \caption{Fit of the transmission spectrum for HD 189733b. From top to bottom: Transmission spectrum of HD 189733b modeled with \texttt{PLATON} with a flat model (red colors) and a model transmission spectrum with a variable Rayleigh-like slope. Residuals after subtraction of the best-fitting model to the flat model and the Rayleigh-like parameterizations.}
    \label{fig:transmission_fit}
\end{figure*}
We perform the first analysis of our retrieved transmission spectrum by fitting a linear model to the transmission planet radii as a function of the logarithmic wavelength, using Eq. \ref{eq:eq1}. The result of the fit and confidence bands can be observed in Fig.\ref{fig:transmission_compare}. We measure a decrease in planet radius of $\Delta R_p / R_{\star}=-0.0082 \pm 0.0013$ along the wavelength range of ESPRESSO, which corresponds to a scattering slope of $-31 \pm 5$ assuming a scale height of $190$ km \citep{Kasper2018ATS}. The slope obtained through this analysis is much higher than values we can find in literature up to date. The reasons behind this excessive slope are partially addressed in \citet{Oshagh2020}, where the authors obtain a lower radius variation computing the RVs with a custom mask combined with the effect of stellar activity. Using only custom CCFs they derive $ \alpha \simeq 9.77\pm 2.72$.In this work, we chose to use the default ESPRESSO masks and try to evaluate the effect of stellar activity afterwards. It is not yet certain what the weight of these effects is and whether ESPRESSO masks suffer from the same problems.\par
To interpret the transmission spectrum, obtained with the CRM technique, we additionally use the forward modeling capabilities of \texttt{PLATON} \citep{Zhang2019, Zhang2020}. We use \texttt{PLATON} assuming an isothermal atmosphere with equilibrium chemistry, divided by default into $500$ layers. The physics of the atmosphere is governed by the hydrostatic equation, setting the planetary radius reference pressure to $1$ bar. In each layer, the code includes the contribution from gas absorption, collisional absorption, and Rayleigh scattering. The contribution of the gas absorption is computed by solving the radiative transfer equation. The presence of different species changes the absorption coefficient of the different layers, changing the opacity as a result. The Rayleigh scattering is controlled by a parametric law, where it is  possible to control the wavelength dependence and the slope strength. In alternative, \texttt{PLATON} supports Mie scattering, which models the contribution of hazes to the transmission spectrum. This model is a function of the imaginary refractive index of the particles, particle size, geometric standard deviation, and fractional scale height. Furthermore, it is possible to select between the already included TiO2, SiO2 amorphous, or solid MgSiO3 particles to model the scattering. \par
For active stars, the presence of unocculted spots and plages is known to change the perceived planet-to-star radius ratio, \citep{2014ApJ...791...55M, 2018ApJ...853..122R} which not only impacts photometric measurements but also RV measurements during transits \citep{Oshagh2014, Boldt2020}. In addition, the inflation does not only affect the white-light radius but can also introduce spurious trends on the transmission spectrum that can mimic a true signal from the planet. To account for this, \texttt{PLATON} introduces a model that is controlled by the fraction of the surface that is plagued with activity and the temperature contrast between these regions and the solar surface. One more additional and important source of attenuation of the spectral features are clouds. With \texttt{PLATON} it is possible to define the pressure of the cloud deck. Below it, the atmosphere is fully opaque and as a consequence, there is no contribution from these layers.\par
We performed several retrievals to understand what is the main mechanism behind the observed slope in the transmission spectrum. The uncertainty was propagated, using a Gaussian prior, for the stellar radius, planetary mass, planet radius, planet temperature, C/O ratio, and atmospheric metallicity. For each of the runs, we computed two distinct inference criteria. $ln(z)$ which corresponds to the natural logarithm of the evidence, that is used to compare directly distinct models, and the $\chi^2$ which is a weighted (by the variance) measure of the distance from the model to the data points, and it can be used to tell us how well the models reproduce the data.\par
We started by constructing a flat model that includes all species and setting the scattering slope to zero. The uncertainties are propagated for the stellar radius, planetary mass ($M_p$), planetary radius, and temperature from the literature using Gaussian priors. In addition, we also propagate in the same way for the log-metallicity (log(Z)), C/O ratio, and temperature of the atmosphere to the values derived in \citet{Zhang2020}. In order to try to reproduce the observed slope in the transmission spectrum, we used the parametric law for the Rayleigh scattering with a uniform prior for the scattering slope and a comprehensive range for the scattering strength $k_{\alpha}$. The priors are summarized in Tab. \ref{table:platon_priors}.\par

\begin{table}[h!]
\centering
\caption{Set of priors for the transmission spectrum fit (R1) with \texttt{PLATON}.}
\label{table:platon_priors}
\begin{tabular}{llll}
\hline
\hline
Parameter & Prior & Posterior\\
\hline
$R_p /R_\star $ & $\mathcal{G}(0.1602, 0.0039)$ & $0.159 \pm 0.014$\\
$R_\star \, (R_{\odot}) $ & $\mathcal{G}(0.766, 0.013)$ & $0.766 \pm 0.010$\\
$M_p \,(M_J)$ & $\mathcal{G}(1.138, 0.025)$ & $1.138 \pm 0.024$\\
$T\,$(K) & $\mathcal{G}(1089, 120)$ & $1079^{+110}_{-106} $ \\
$| \alpha |$ & $\mathcal{U}(2, 50)$ & $22^{+7}_{-6} $\\
$log(k_{\alpha})$ & $\mathcal{U}(-4, 4)$& $-0.43^{+2.01}_{-2.12} $\\
log(Z) & $\mathcal{G}(1.08, 0.23)$ & $1.06 \pm 0.22$\\
C/O & $\mathcal{G}(0.66,0.09)$ & $0.66 \pm 0.09$\\
\hline
\end{tabular}
\tablefoot{The reference parameters are from \citet{Zhang2020}, and the planet-to-star radius ratio results from the white-light fit from the ESPRESSO data. In the last column, the fit median values and uncertainties.}
\end{table}
Fig. \ref{fig:transmission_fit} presents our results for the models with and without scattering. For the flat model (R0) we obtain a log evidence $ln(B_{R0})=154.06 \pm 0.10$ and a $\chi ^2  = 63.76$. We retrieve the best-fit model with the parametric scattering (R1) with $ln(B_{R1})=160.07\pm 0.11$ and $\chi ^2  = 33.68$. The Bayes factor between this model and the flat one is $407 \pm 61$, which represents very strong evidence against R0 \citep{Kass95bayesfactors}. The model seems also to better reproduce the global trend of the data, especially in the bluest range where it is able to fit the small plateau that is observed. However, it is unable to explain the radius measurements centered at $468.975$ nm and $680.295$ nm which are of by over 2-sigma. This model produces a decrease in radius over the entire range of wavelengths of $\Delta R_p / R_{\star}=-0.0046 \pm 0.0008$, which is significantly different from the value we derived before. Increasing the number of species in the model seems to explain, in part, the observed slope and produces a significantly lower $\alpha \simeq 22.4 \pm 7$.\par
\section{Conclusions}
In this paper, we used the chromatic Rossiter-McLaughlin effect applied to high-resolution ESPRESSO data to retrieve the transmission spectrum of HD 189733b.\par
We started by fitting the white-light RVs with a model composed of a Keplerian component and the effect of a transiting exoplanet on the integrated CCF.\par
We tentatively detect the presence of differential rotation with a confidence of $93.4\%$ and derive an equatorial rotation period of $11.45 \pm 0.09$ days and a polar period of $14.9 \pm 2$ days. Additionally, using a first-degree surface brightness (CB) model, we derive a convective blueshift scale factor of $S \approx 0.60^{+0.20}_{-0.18}$. We also test a broader range for the rotation period, which increases the confidence for the differential rotation ratio being larger than $0.05$ to $99.6\%$. The median value for the convective blueshift is compatible in both scenarios. \par
The presence of differential rotation breaks the symmetry of the stellar RV field, which further allows the determination of the stellar tilt. We find $i_{\star} \approx 71.87^{+6.91^\circ}_{-5.55^\circ}$ and a projected spin-orbit angle of $\lambda \approx -1.00^{+0.22^\circ}_{-0.23^\circ}$. In turn, we compute the true 3D spin-orbit angle ($\psi \approx 13.6 \pm 6.9^\circ $) and note that the planet seems to be well aligned.\par

We analyze the transmission spectrum by first fitting a simple Rayleigh scattering model to the data. We find a greatly enhanced Rayleigh scattering slope, often called Super-Rayleigh, of $-31 \pm 5$ and a radius decrease of $\Delta R_p / R_{\star}=-0.0082 \pm 0.0013$. Using the forward retrieval software \texttt{PLATON}, with all the atomic and molecular species available, we estimate a significantly lower slope ($\alpha \simeq -22.4 \pm 7$) and radius variation: $\Delta R_p / R_{\star}=-0.0046 \pm 0.0008$. We reproduce the plateau observed in the blue range ($378.05-443.26$ nm), which cannot be explained by scattering alone. The enhanced radius-wavelength slope has been explored in the literature, with some authors being able to emulate the decrease using models of the stellar surface with unocculted cold spots \citep[e.g.,][]{2014ApJ...791...55M, Kasper2018ATS, 2019AJ....157...96R} or occultation of plages \citep{Boldt2020}. However the origin of the stronger slope observed in ground-based observations requires further investigation. It is plausible that STIS observations may be less sensitive to stellar activity, as they primarily rely on variations in flux contrast on the stellar surface during transits. Additionally, spectroscopic observations can exhibit line profile deformations induced by stellar activity, potentially varying with wavelength. This characteristic could be specific to active stars since a similar approach was used in \citep{Santos2020} to retrieve the broadband transmission spectrum of HD 209458b did not result in any detectable strong slope in the blue wavelengths. To understand the origin of these differences and determine the more reliable method, it is crucial to conduct high precision simultaneous observations in the future.

\begin{acknowledgements}
   The authors would like to express their gratitude to the ESPRESSO project team for their effort and dedication in building the ESPRESSO instrument.

   EC was supported by Fundação para a Ciência e a Tecnologia (FCT, Portugal) through the research grants UIDB/04434/2020, UIDP/04434/2020, and PRT/BD/152703/2022.
   
   The authors would also like to acknowledge the invaluable comments from the anonymous referee, which greatly helped improve the clarity of ideas and enhance the quality of this work.
   
   The research leading to these results has received funding from the European Research Council through the grant agreement 101052347 (FIERCE).
   
   This work was supported by FCT - Fundação para a Ciência e a Tecnologia through national funds and by FEDER through COMPETE2020 - Programa Operacional Competitividade e Internacionalização with the grants UIDB/04434/2020 and UIDP/04434/2020.
   
   JIGH, RR, CAP, and ASM acknowledge financial support from the Spanish Ministry of Science and Innovation (MICINN) project PID2020-117493GB-I00.
   
   ASM, JIGH, and RR also acknowledge financial support from the Government of the Canary Islands project ProID2020010129.
   
   ASM acknowledges financial support from the Spanish Ministry of Science and Innovation (MICINN) under the 2018 Juan de la Cierva program IJC2018-035229-I.
   
   CJM acknowledges FCT and POCH/FSE (EC) support through Investigador FCT Contract 2021.01214.CEECIND/CP1658/CT0001.
   
   NJN acknowledges financial support by FCT - Fundação para a Ciência e a Tecnologia under projects UIDB/04434/2020 and UIDP/04434/2020, CERN/FIS-PAR/0037/2019, and PTDC/FIS-AST/0054/2021.
   
   MRZO acknowledges financial support from the Spanish Ministry of Research and Innovation through project PID2019109522GB-C51.
   
   This project has received funding from the European Research Council (ERC) under the European Union's Horizon 2020 research and innovation programme (project "Spice Dune," grant agreement No 947634).
   
   J.L-B. acknowledges financial support received from "la Caixa" Foundation (ID 100010434) and from the European Union Horizon 2020 research and innovation programme under the Marie Skłodowska-Curie grant agreement No 847648, with fellowship code LCF/BQ/PI20/11760023.
   
   This research has also been partly funded by the Spanish State Research Agency (AEI) Projects No. PID2019-107061GB-C61.
   
   The contributions of AP and ML have been carried out within the framework of the NCCR PlanetS supported by the Swiss National Science Foundation under grants 51NF40-182901 and 51NF40-205606. ML acknowledges support from the Swiss National Science Foundation under grant number PCEFP2-194576.

\end{acknowledgements}
 
\bibliographystyle{aa.bst}
\bibliography{crm.bib}
\begin{appendix}\label{appendix}

\onecolumn{
\section{White-light corner plot for M1}\label{a_white_corner}
\begin{figure}[h!]
    \centering
    \includegraphics[width=\linewidth]{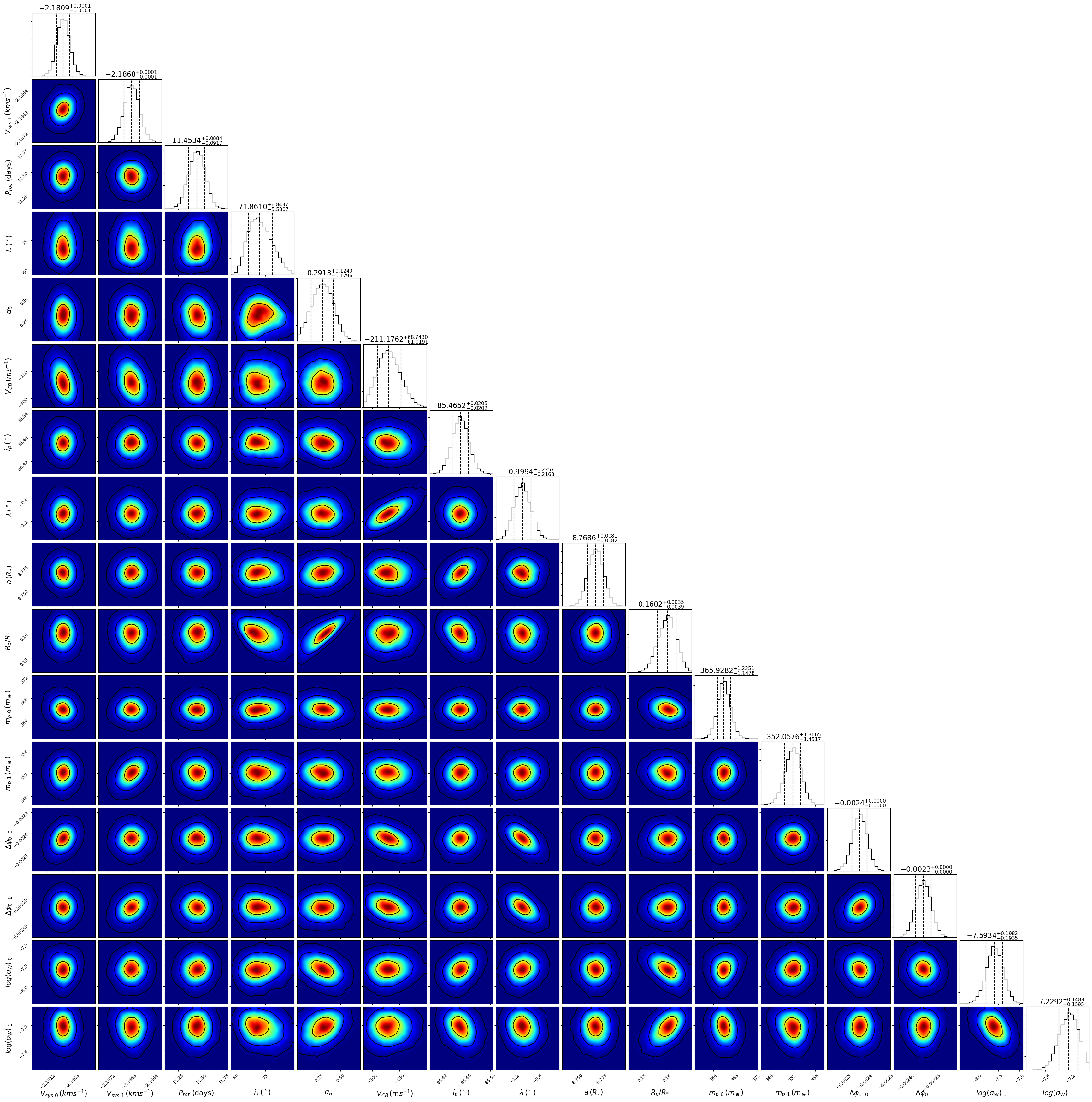}
    \caption{Corner plot of the joint white-light fit of the two observations of HD 189733 for M1. The values and errors correspond to the median and the standard deviation of the posterior distributions (assuming they are normal). The subscript numbers in the variables $V_{sys}$, $\Delta \phi_0$, $m_P$ and $\log(sigma_W)$ represent the posterior distribution of the parameters that were independently fitted for the two nights.}
    \label{fig:corner_white}
\end{figure}
}
\onecolumn{
\section{White-light corner plot for M2}\label{a_white_corner_M2}
\begin{figure}[h!]
    \centering
    \includegraphics[width=\linewidth]{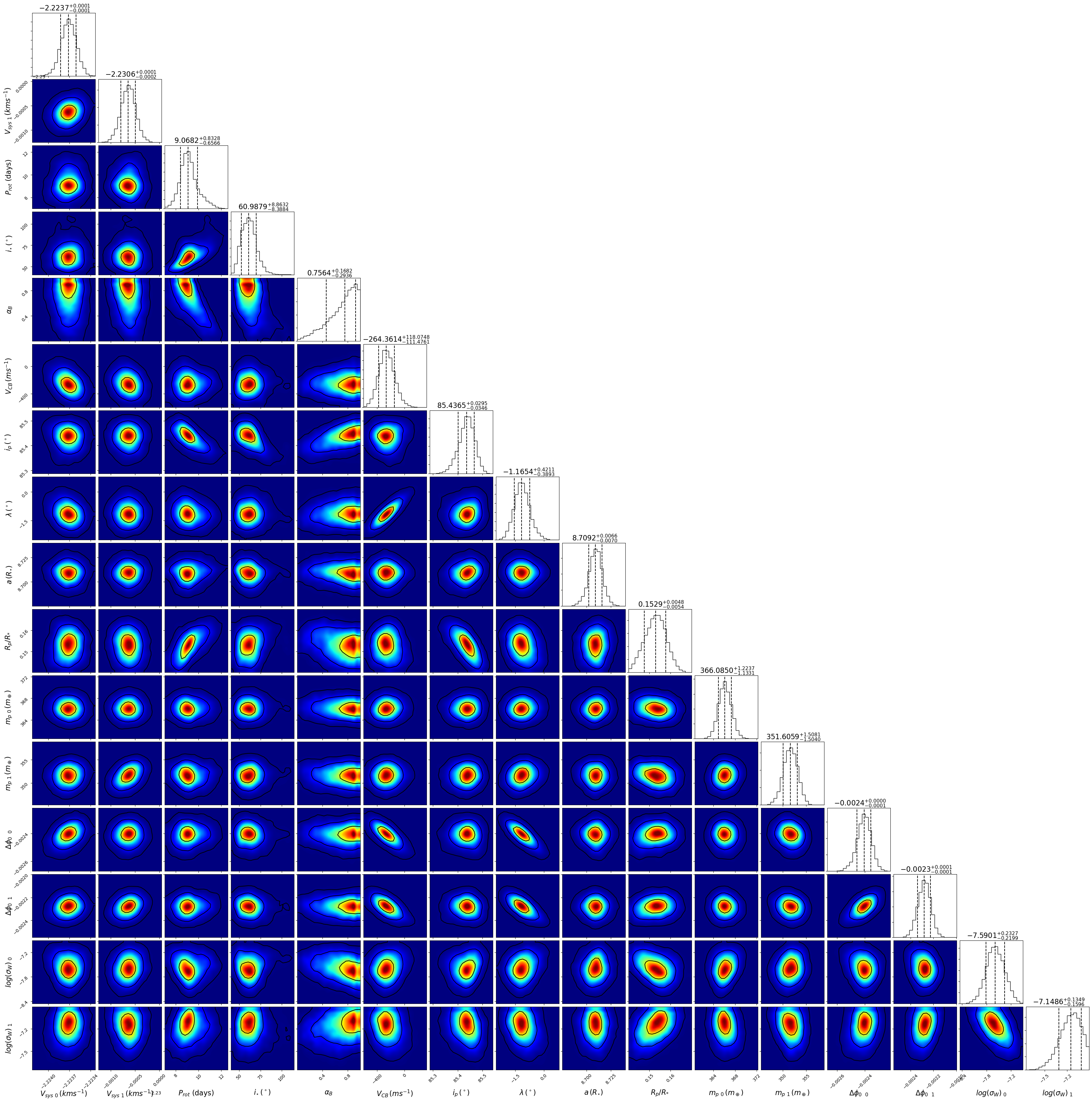}
    \caption{Corner plot of the joint white-light fit of the two observations of HD 189733 for M2. The values and errors correspond to the median and the standard deviation of the posterior distributions (assuming they are normal). The subscript numbers in the variables $V_{sys}$, $\Delta \phi_0$, $m_P$ and $\log(sigma_W)$ represent the posterior distribution of the parameters that were independently fitted for the two nights.}
    \label{fig:corner_white_m2}
\end{figure}
}
\onecolumn{
\section{Individual night fits}\label{a_individual_fits}
\begin{figure}[ht!]
\centering
  \includegraphics[width=0.49\linewidth]{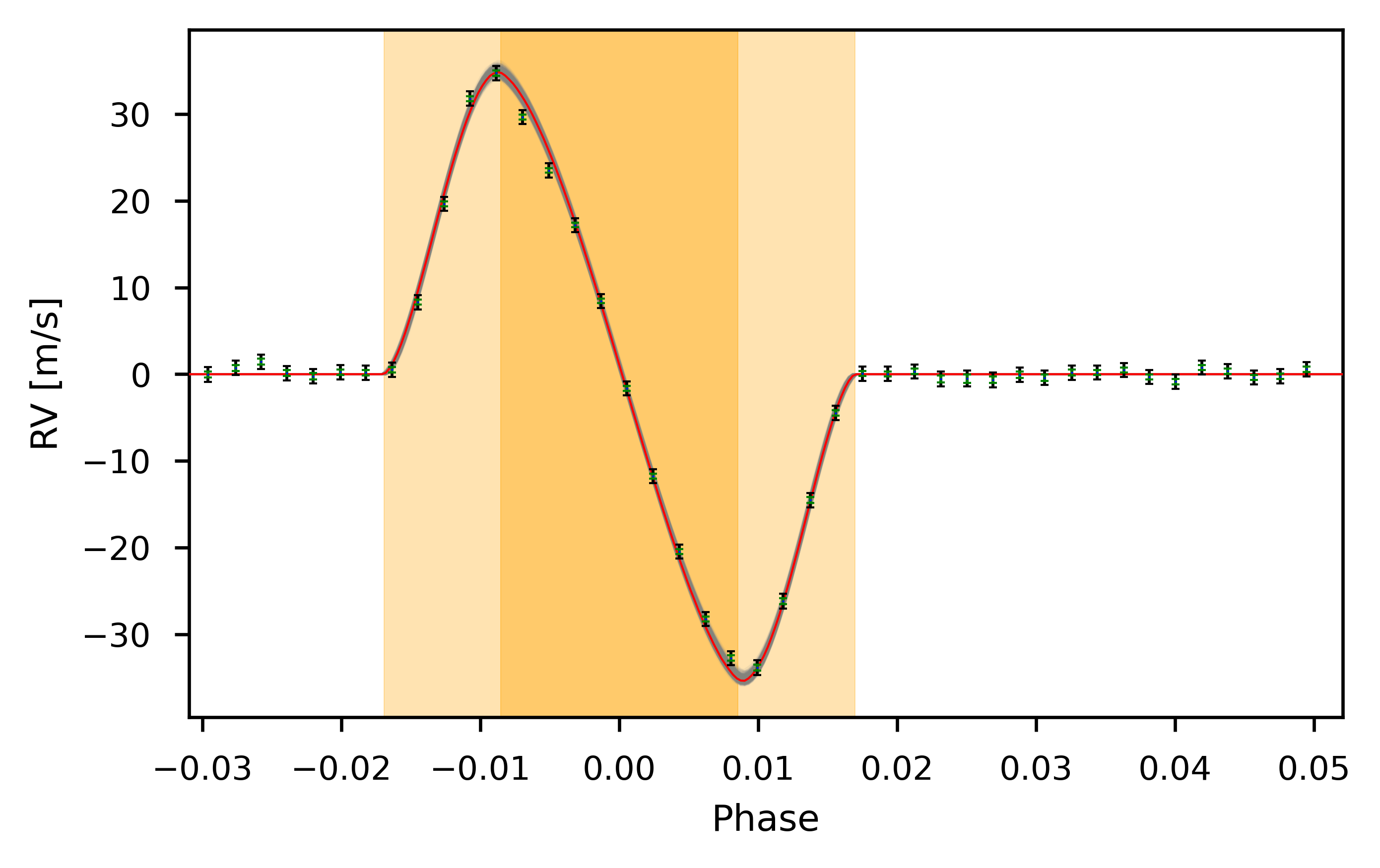}
 \includegraphics[width=.49\linewidth]{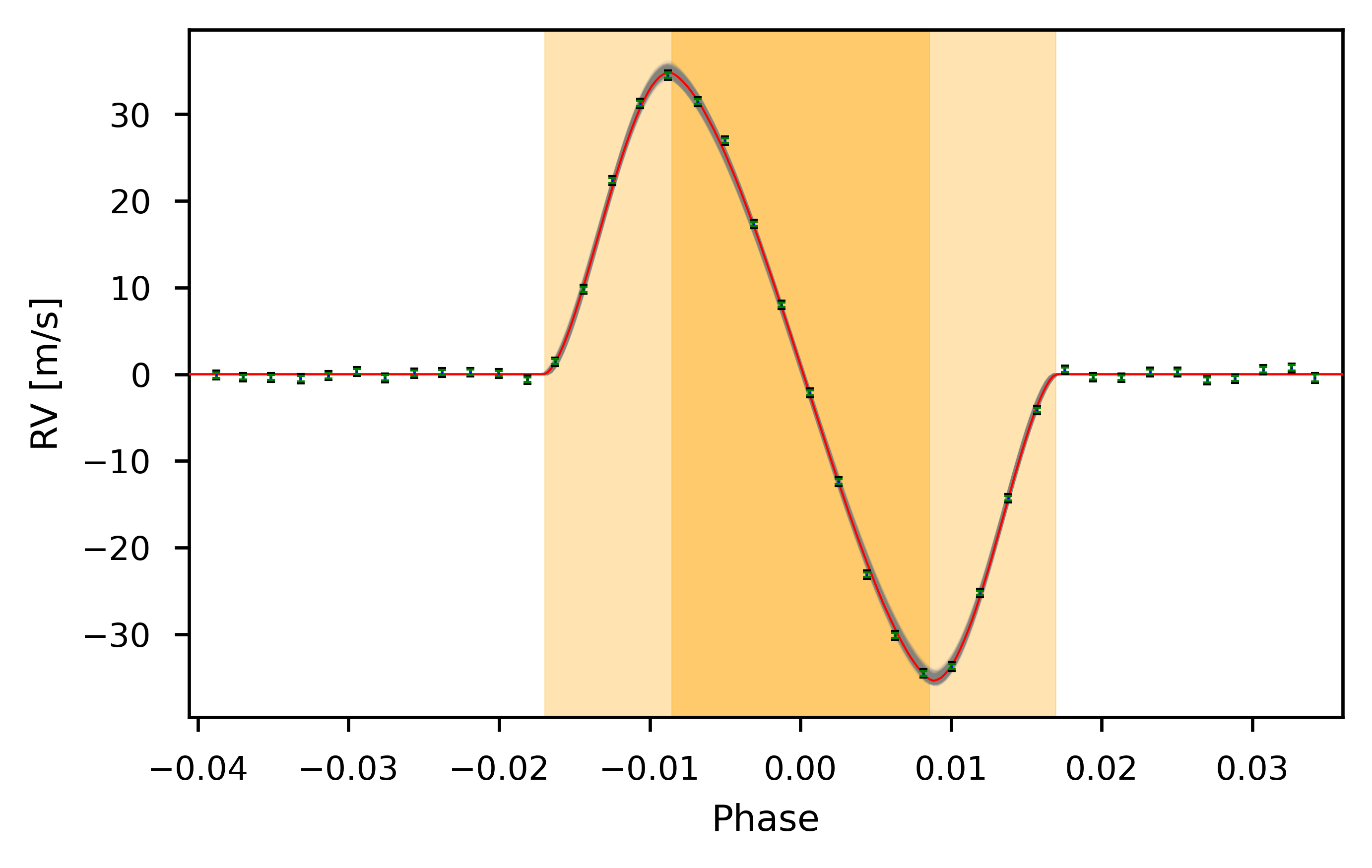}
   \includegraphics[width=0.49\linewidth]{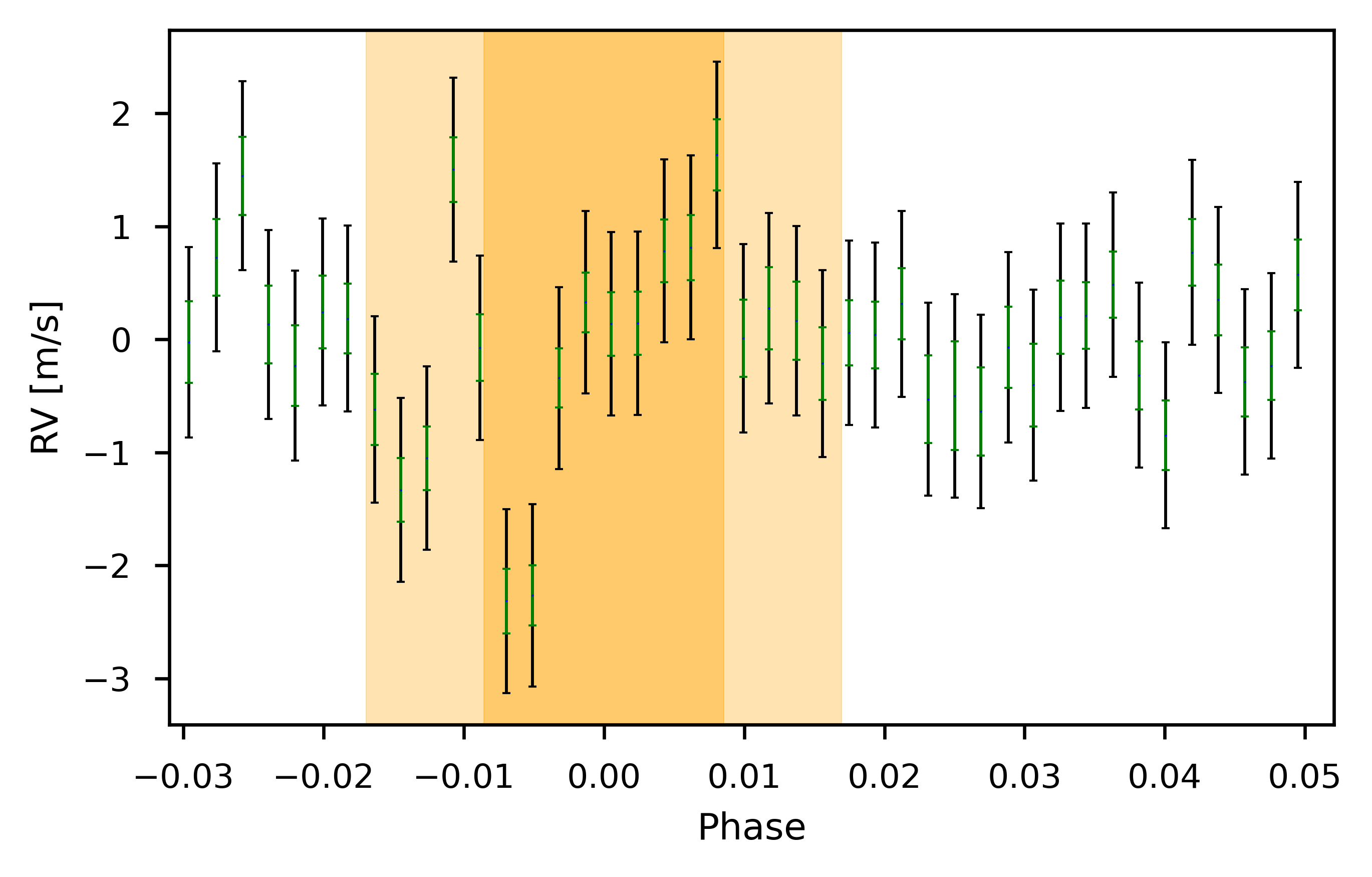}
  \includegraphics[width=.49\linewidth]{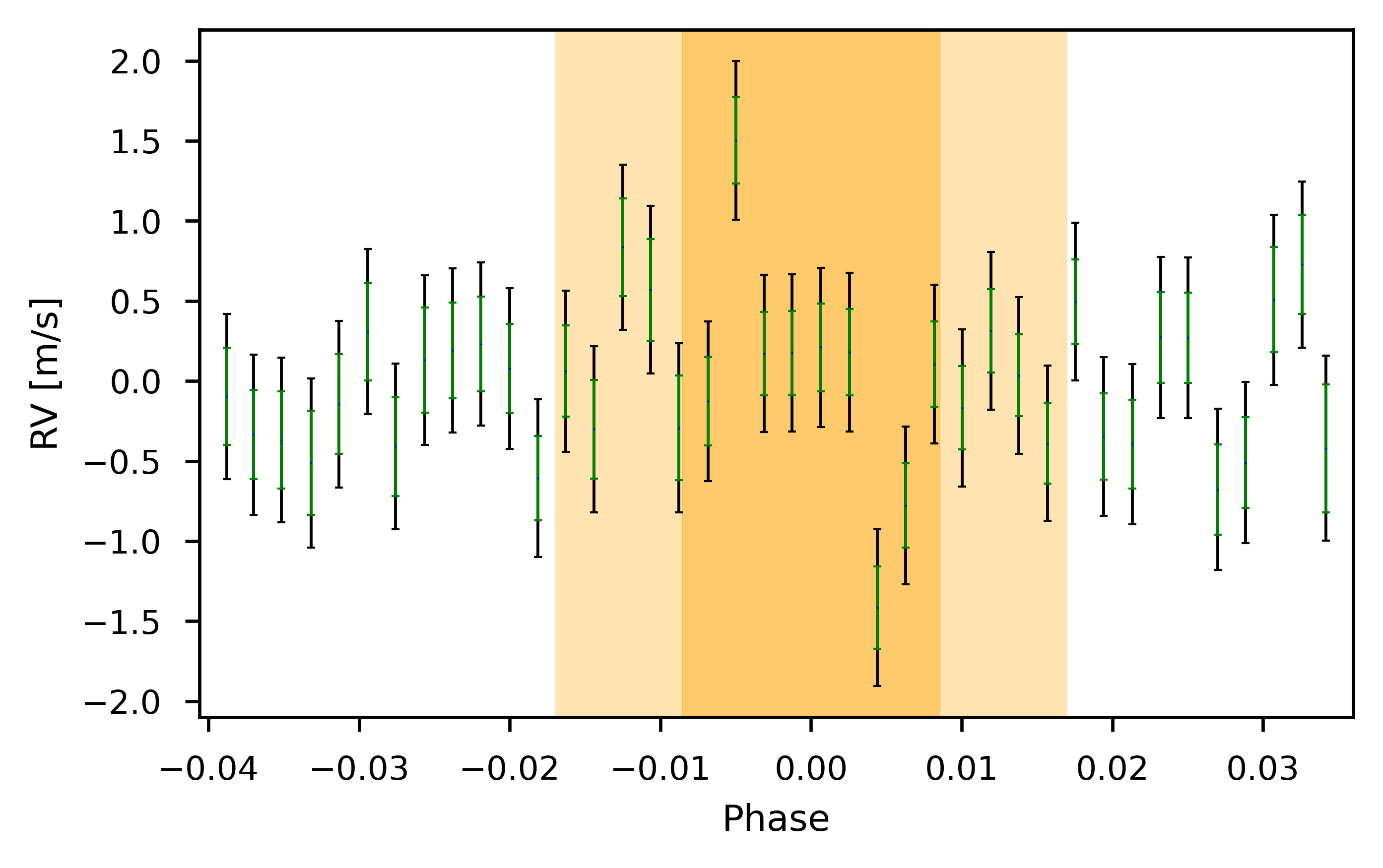}
    \includegraphics[width=.49\linewidth]{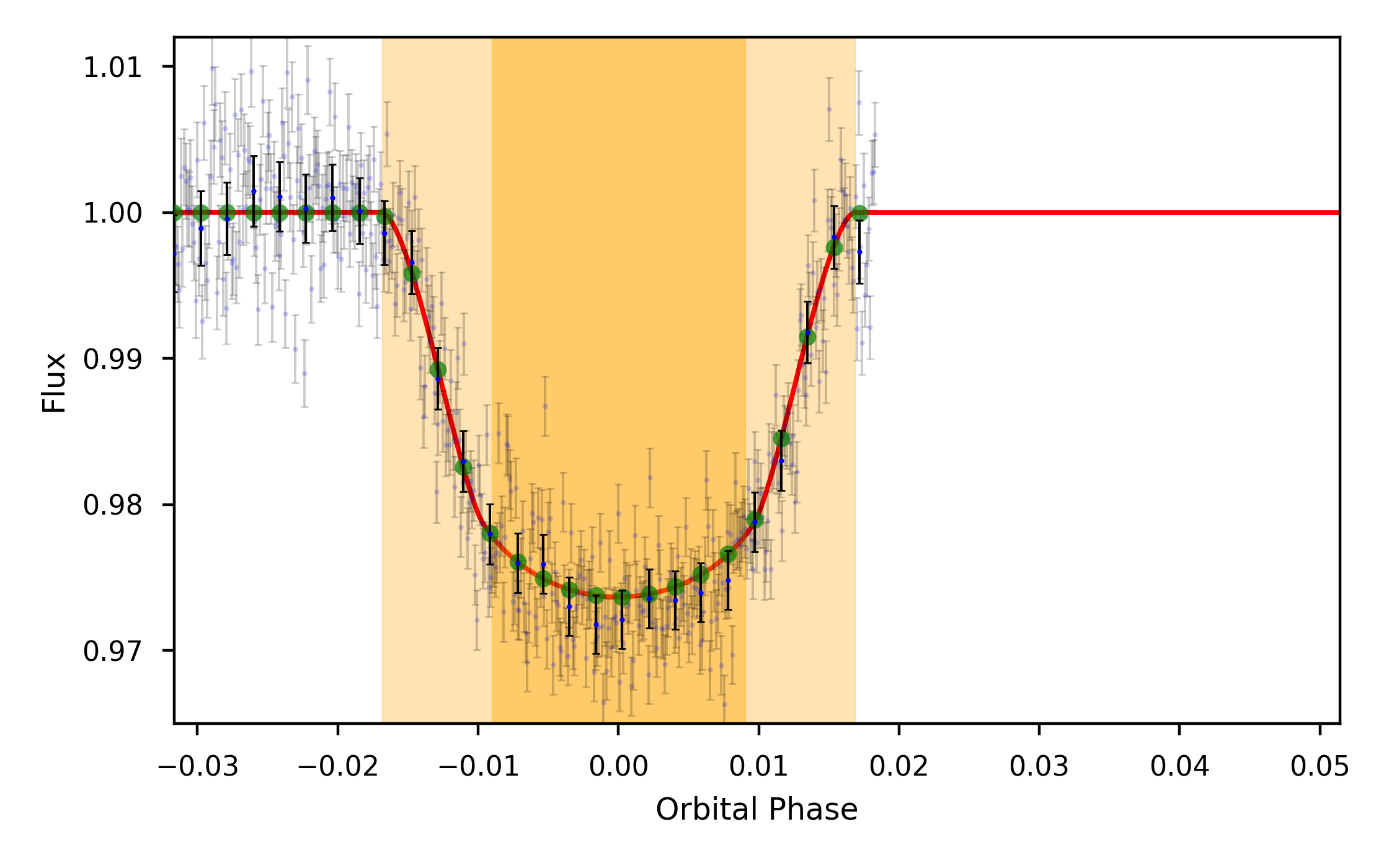}
    \includegraphics[width=.49\linewidth]{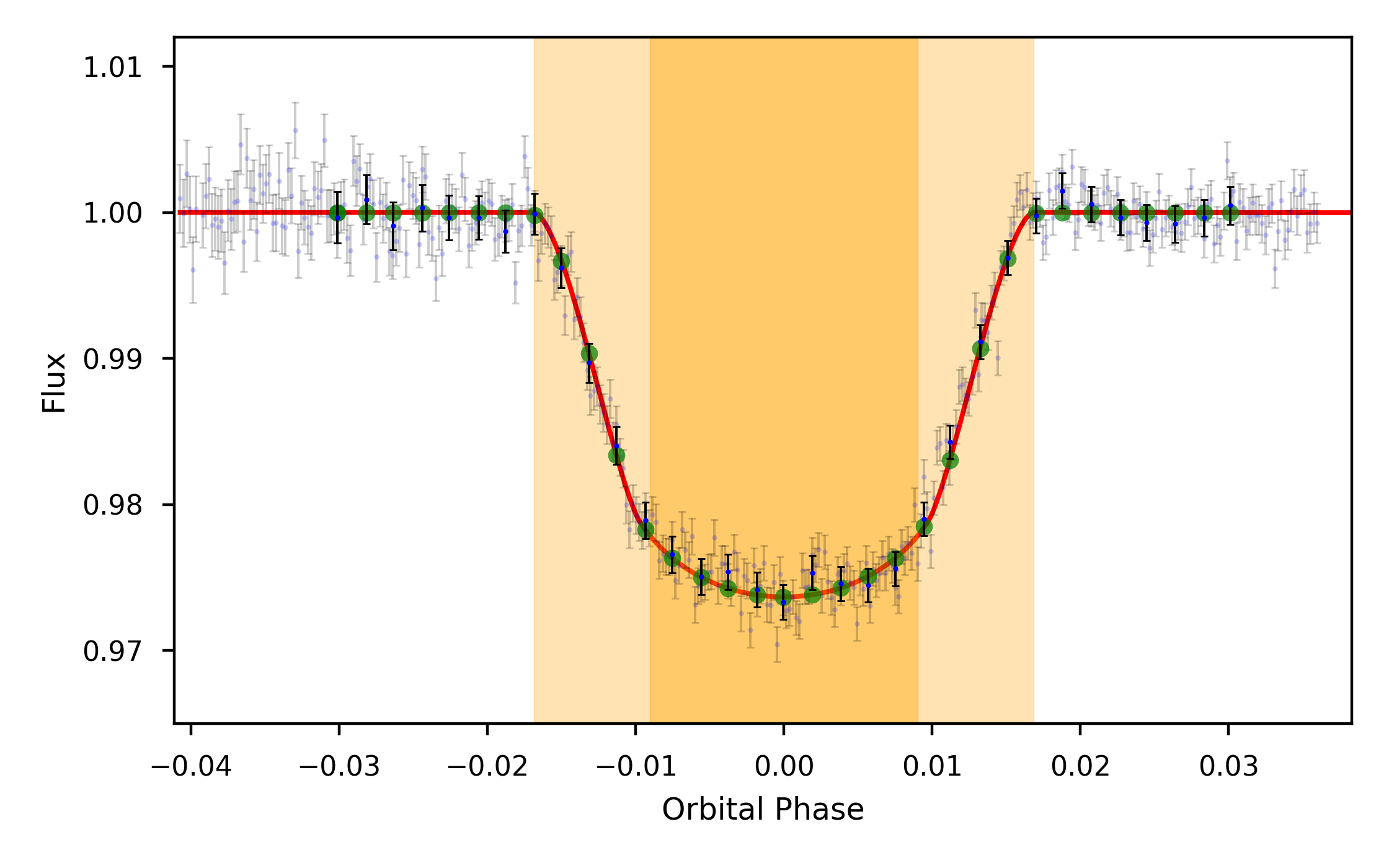}
        \includegraphics[width=.49\linewidth]{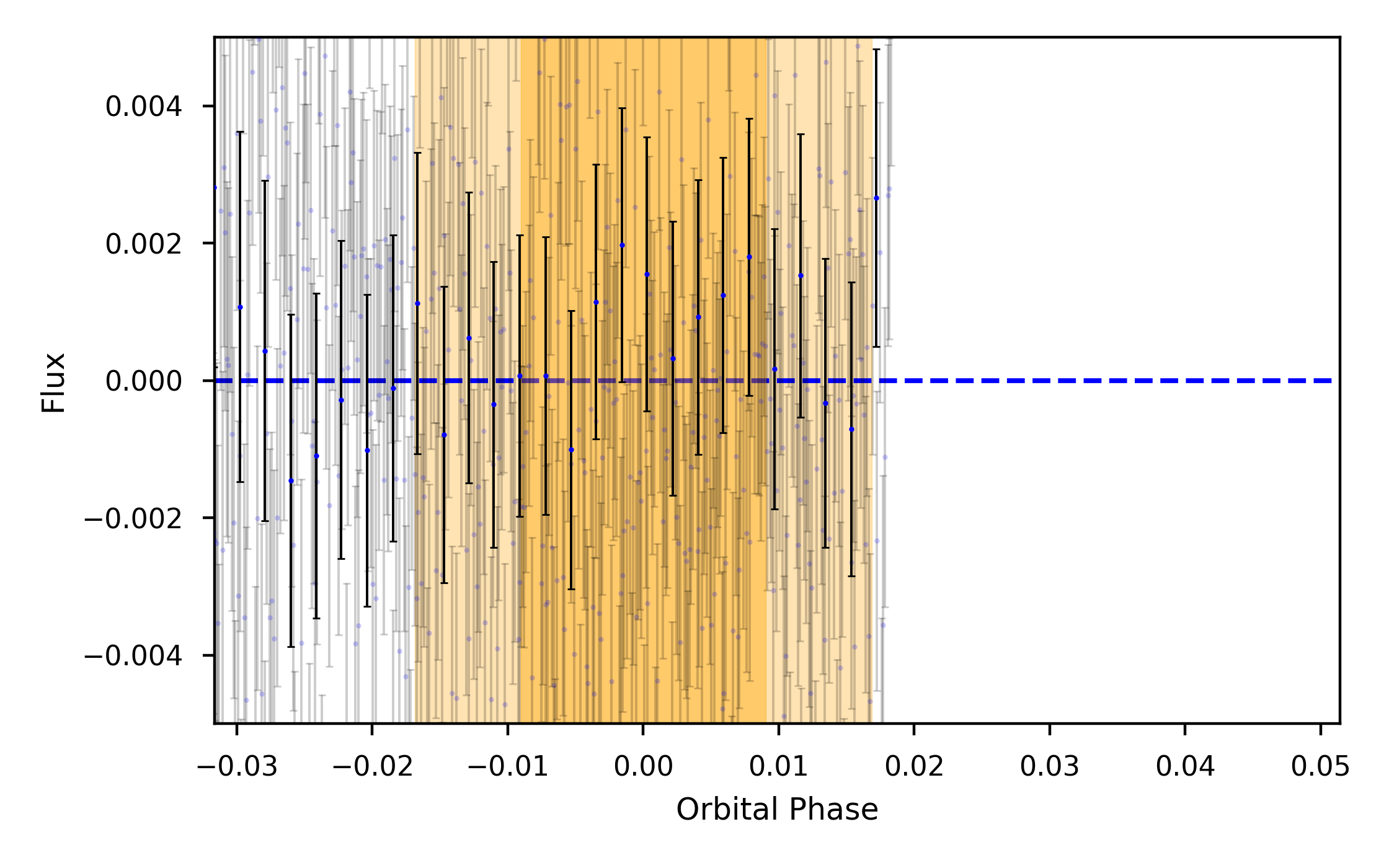}
    \includegraphics[width=.49\linewidth]{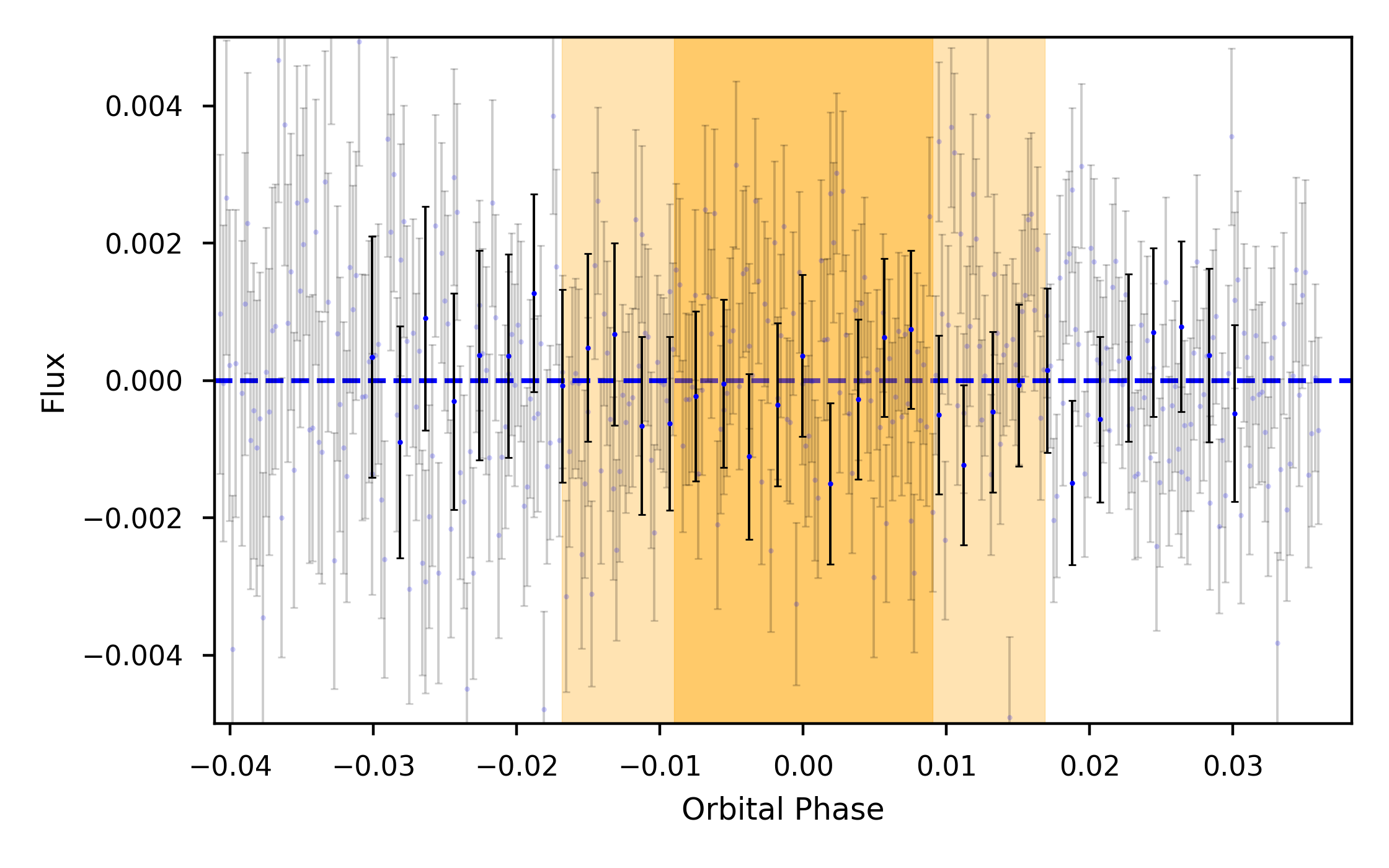}

  \caption{Data fit and residuals for the individual nights. Same as Fig. \ref{fig:hd189jres} for the individual nights.}
\label{fig:hd189_individual}
\end{figure}
}

\onecolumn{
\section{Doppler Maps of the stellar surface and the RM components}\label{Doppler Maps}
\begin{figure}[ht!]
    \centering
    \includegraphics[width=.615\linewidth]{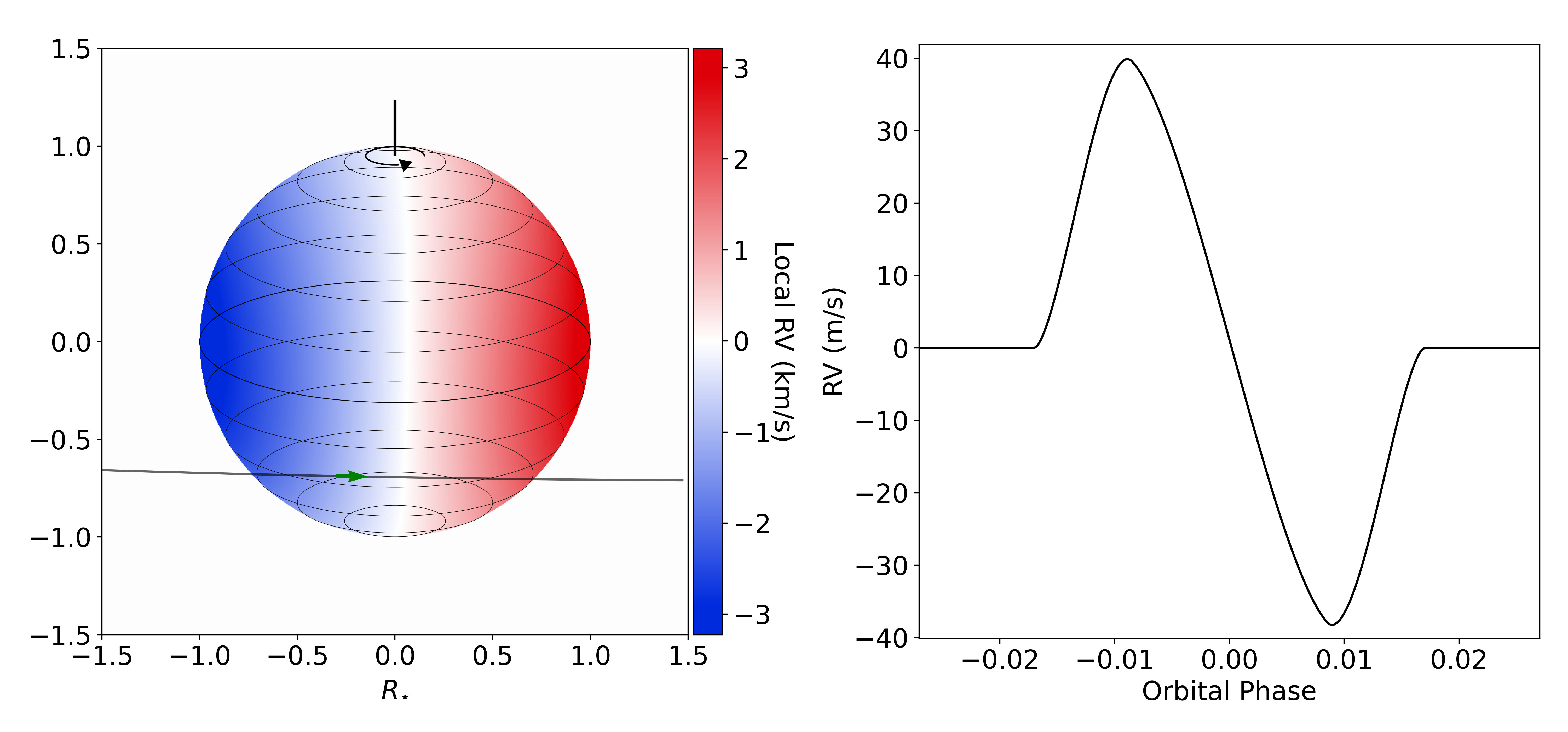}
    \includegraphics[width=.615\linewidth]{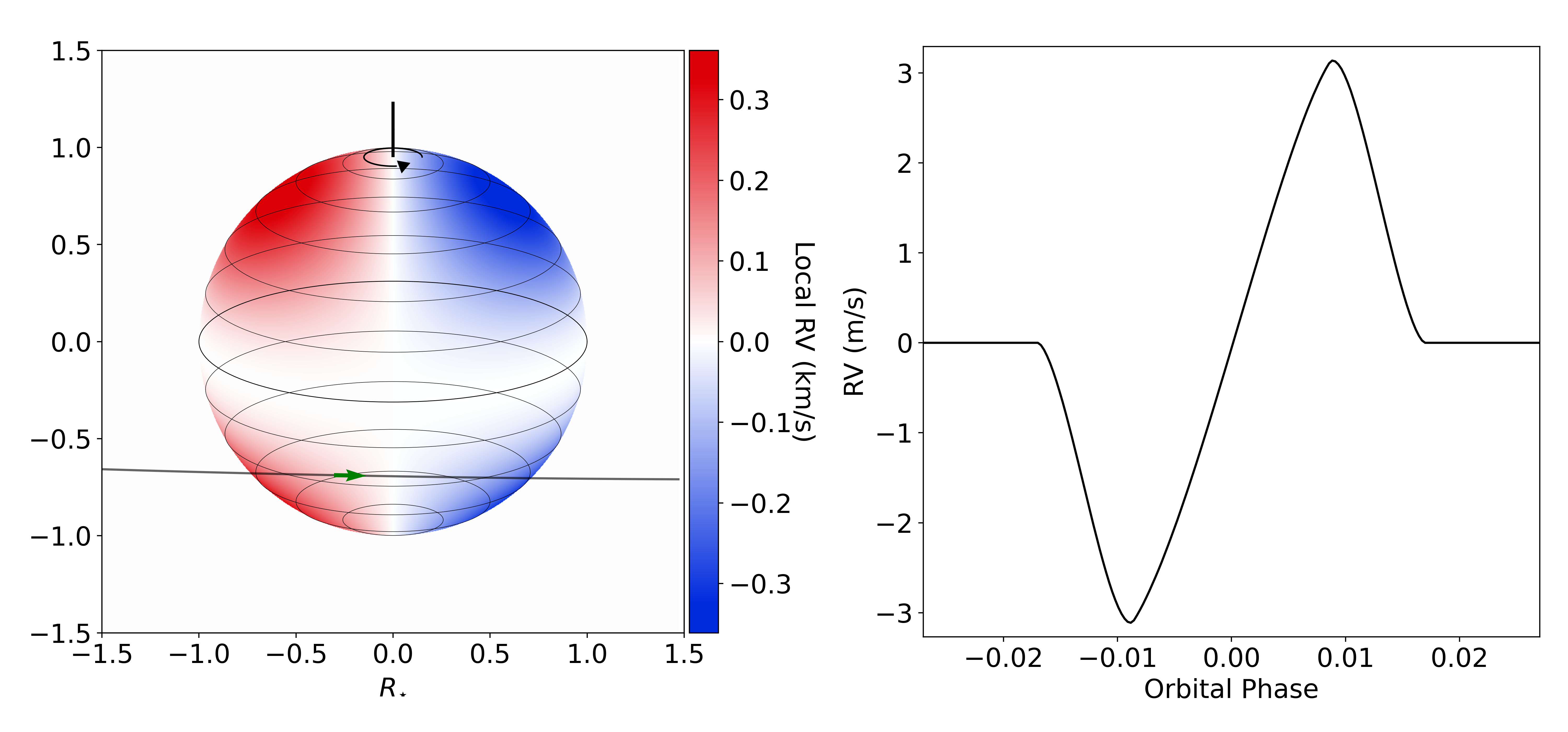}
    \includegraphics[width=.615\linewidth]{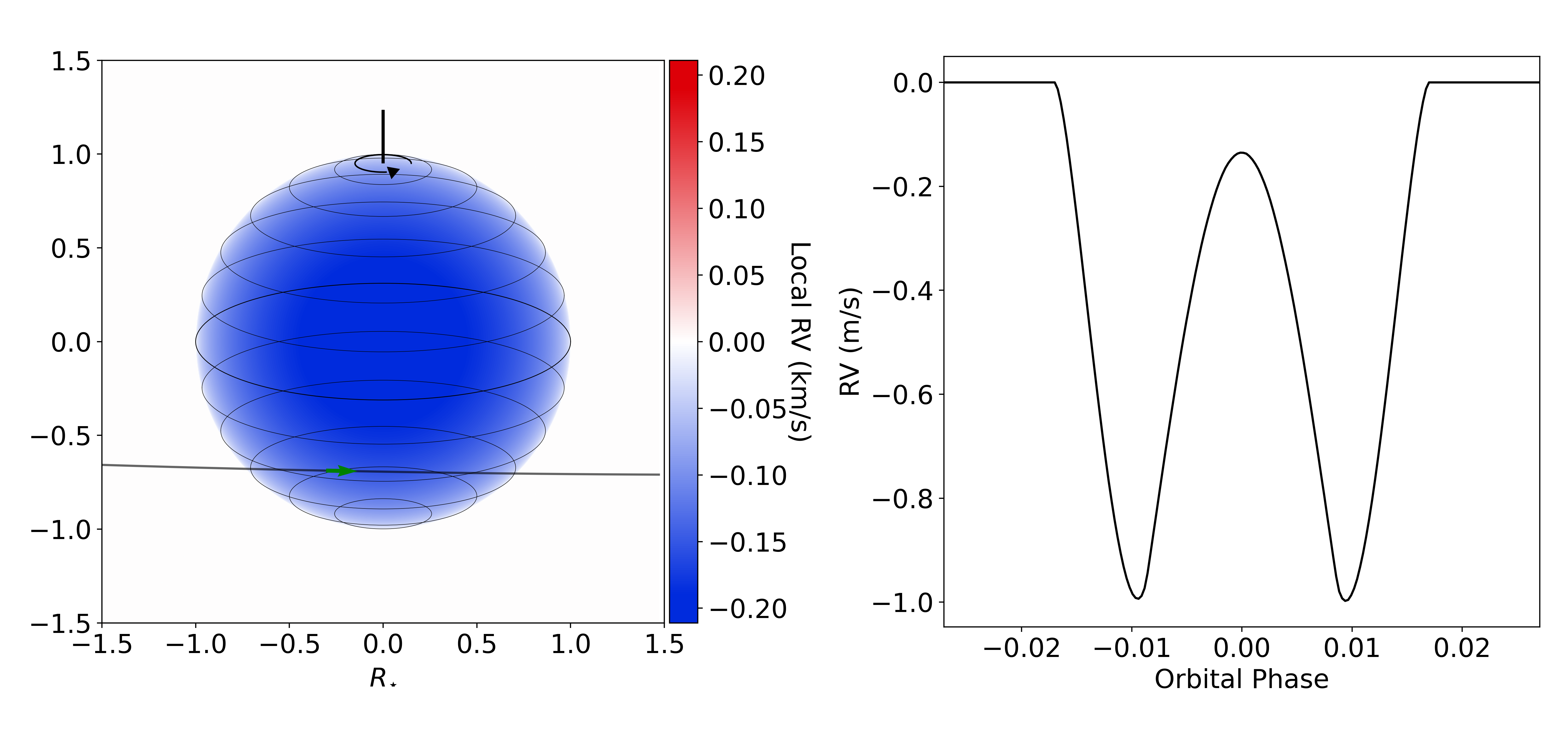}
    \includegraphics[width=.615\linewidth]{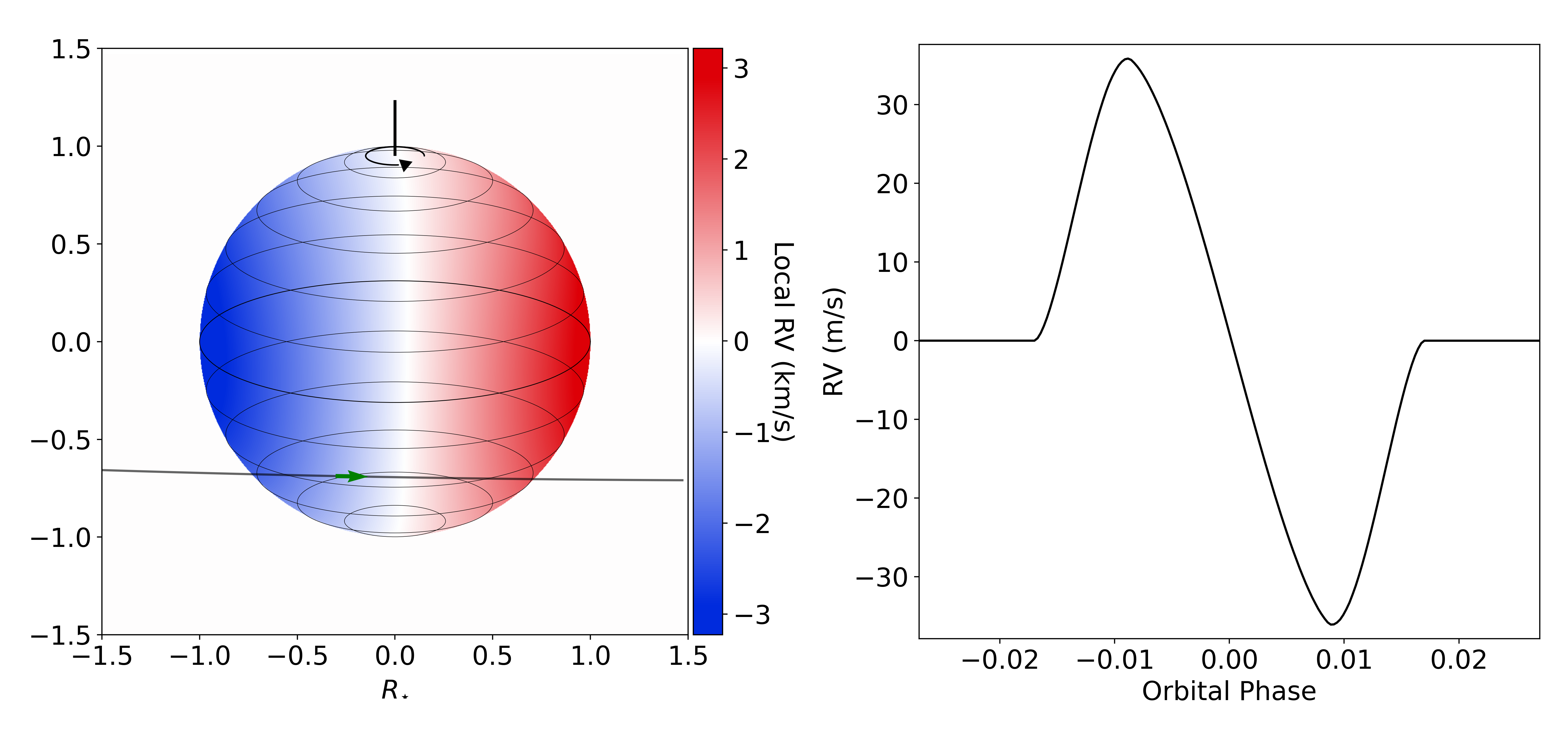}

    \caption{Decomposition of the Doppler surface maps and corresponding RV anomaly produced by the transiting exoplanet. Left: Doppler maps of the stellar surface for the best-fit solution for the HD 189733 data. The black lines over the maps show the latitudinal lines in intervals of 15$^\circ$.  The curved black and straight green arrows indicate the rotation direction of the star and the planet along its orbits, respectively. Right: RV anomaly created by the transiting exoplanet for the corresponding (same row) Doppler map. From top to bottom: rigid-body rotation model for the equatorial stellar velocity, difference after subtracting the rigid-body rotation model from the differential rotation model, convective blueshift, and sum of all the previously mentioned components.}
    \label{fig:dopplermap}
\end{figure}

}
\begin{figure}
    \centering
    \includegraphics[width=\linewidth]{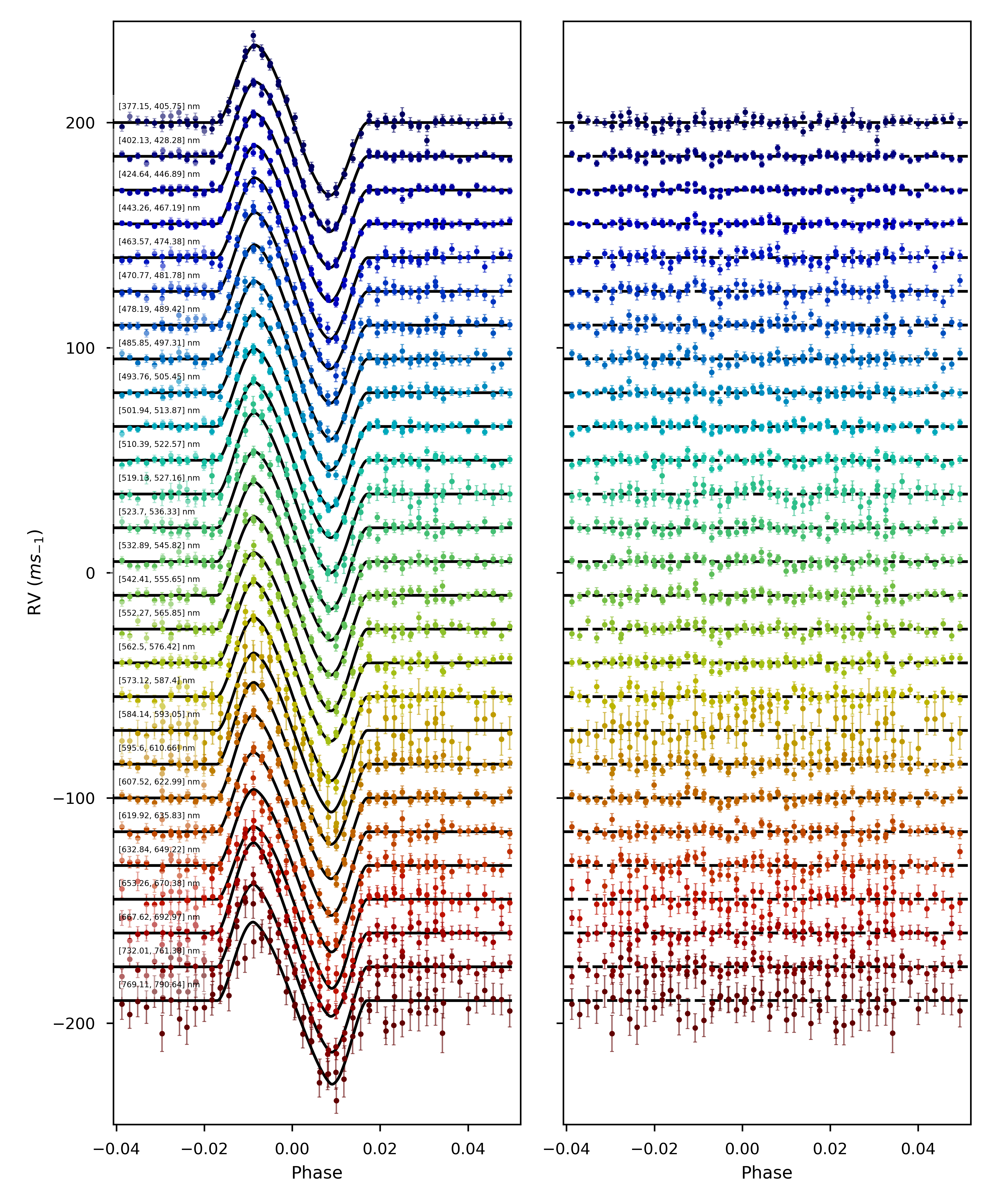}
    \caption{Chromatic RM fits and corresponding residuals. Left: The RVs computed for the chromatic ranges and the best-fit model (black line). Right: Residuals after subtraction to the data of the model corresponding to the best solution. The RV data, models, and corresponding residuals are shifted by $15 m\,s^{-1}$ increments for better visualization.}
    \label{fig:chromatic_rm_fits}
\end{figure}

\twocolumn

\section{Reference values table}
\begin{table}[h!]
\centering
\caption{Reference values for the retrievals with \texttt{CaRM} using the \texttt{SOAP} model. }
\label{table:tablepars}
\begin{tabular}{lll}
\multirow{2}{*}{} & \multicolumn{2}{l}{\textbf{HD189733}} \\
 \hline
Parameter & Value & Source \\
\hline
$R_\star\,(R_{\rm \odot})$      & $0.766^{+0.007}_{-0.013}$                 & \cite{Triaud2009} \\
$T_{\rm eff} \,$(K)             & $4969 \pm 43$ 			                & \cite{Sousa2018} \\
$\log(g)$                       & $4.60 \pm 0.01$ 						    & \cite{Sousa2018}  \\
$\rm [Fe/H]$ (dex)	            & $-0.07 \pm 0.02$						    & \cite{Sousa2018} 	\\
$T_0 \,$(MBJD) 		            & $53988.30339 ^{+0.000072}_{-0.000039}$ 	& \cite{Triaud2009}\\
$P \,$(days) 		            & $2.21857312 ^{+0.00000036}_{-0.00000076}$ & \cite{Triaud2009} \\
$a \,(R_\star)$ 		        & $8.756 ^{+0.0092}_{-0.0056}$ 			    & \cite{Triaud2009} \\
$\lambda \,(^\circ)$ 		    & $-0.85^{+0.28}_{-0.32}$ 				    & \cite{Triaud2009} \\
$P_{rot}$ (days)      & $11.953 \pm 0.009$						& \cite{Henry_2007}\\
$R_{\rm p} \,(R_\star)$ 		& $0.1581 \pm 0.0005$ 				        & \cite{Triaud2009}\\
$K \,$(m\,s$^{-1}$) 			    & $ 201.96 ^{+1.07}_{-0.63} $ 		        & \cite{Triaud2009}\\
$i_p \,(^\circ)$ 					& $85.508 ^{+0.10}_{-0.05}$ 				& \cite{Triaud2009}\\
\hline
\end{tabular}
\end{table}
\section{Transmission spectrum}
\begin{table}[h!]
    \centering
    \caption{Planet radius, obtained by our analysis,
as a function of wavelength.}
    \label{tab:radius_table}
    \begin{tabular}{cccc}
        \hline
        \hline
        $\lambda$ (nm) & $\Delta \lambda$ (nm) & $R_p/R_*$ & $\Delta R_p/R_*$ (upper/lower)\\
        \hline
        391.350 & 13.300 & 0.16503 & 0.00134 / 0.00132 \\
        415.205 & 13.075 & 0.16481 & 0.00086 / 0.00093 \\
        435.765 & 11.125 & 0.16507 & 0.00070 / 0.00068 \\
        455.225 & 11.965 & 0.16593 & 0.00074 / 0.00076 \\
        468.975 & 5.405  & 0.16832 & 0.00106 / 0.00124 \\
        476.275 & 5.505  & 0.16548 & 0.00118 / 0.00119 \\
        483.805 & 5.615  & 0.16529 & 0.00125 / 0.00126 \\
        491.580 & 5.730  & 0.16487 & 0.00113 / 0.00111 \\
        499.605 & 5.845  & 0.16404 & 0.00097 / 0.00093 \\
        507.905 & 5.965  & 0.16484 & 0.00079 / 0.00092 \\ 
        516.480 & 6.090  & 0.16220 & 0.00111 / 0.00129 \\
        523.145 & 4.015  & 0.16466 & 0.00166 / 0.00156 \\
        530.015 & 6.315  & 0.16288 & 0.00130 / 0.00134 \\
        539.355 & 6.465  & 0.16264 & 0.00120 / 0.00112 \\
        549.030 & 6.620  & 0.16267 & 0.00108 / 0.00102 \\
        559.060 & 6.790  & 0.16219 & 0.00134 / 0.00123 \\
        569.460 & 6.960  & 0.16238 & 0.00108 / 0.00106 \\
        580.260 & 7.140  & 0.16428 & 0.00134 / 0.00142 \\
        588.595 & 4.455  & 0.16139 & 0.00358 / 0.00362 \\
        603.130 & 7.530  & 0.16268 & 0.00133 / 0.00137 \\
        615.255 & 7.735  & 0.16319 & 0.00115 / 0.00116 \\
        627.875 & 7.955  & 0.16238 & 0.00128 / 0.00131 \\
        641.030 & 8.190  & 0.16170 & 0.00157 / 0.00164 \\
        661.820 & 8.560  & 0.16054 & 0.00244 / 0.00243 \\
        680.295 & 12.675 & 0.16684 & 0.00152 / 0.00175 \\
        746.695 & 14.685 & 0.16211 & 0.00186 / 0.00188 \\
        779.875 & 10.765 & 0.15820 & 0.00377 / 0.00388 \\
        \hline
         
    \end{tabular}
\tablefoot{ The $\lambda$ symbol corresponds to the central wavelength and $\Delta \lambda$ the bin width. The uncertainties in the radius correspond to a 68$\%$ confidence interval, with the upper and lower intervals around the median value.}
\end{table}

\end{appendix}
\end{document}